\documentclass[aip,jcp,reprint,showkeys]{revtex4-1}

\usepackage[version=3]{mhchem} % Formula subscripts using \ce{}
\usepackage{graphicx}
\usepackage{multirow}
\usepackage{amsmath}
\usepackage{amssymb}
\usepackage{nicefrac}
\usepackage{booktabs}
\usepackage{subcaption}
\usepackage{mathtools}
\usepackage{mwe}
\usepackage{array}
\newcolumntype{m}{>{$} c <{$}}
\usepackage{color,amscd,amsmath,amssymb,amsfonts,physics,siunitx}

\def\rv{{\bf r}}
\def\Rv{{\bf R}}

\def\tv{{\bf t}}

\def\dd{\mathrm{d}}

\def\xv{{\bf x}}
\def\beq{\begin{equation}}
\def\eeq{\end{equation}}

\def\erf{\mathrm{erf}}
\def\erfc{\mathrm{erfc}}
\def\loc{\mathcal{L}}

\begin{document}     

\author{Kimberly J. Daas}
\affiliation
{Department of Chemistry \& Pharmaceutical Sciences and Amsterdam Institute of Molecular and Life Sciences (AIMMS), Faculty of Science, Vrije Universiteit, De Boelelaan 1083, 1081HV Amsterdam, The Netherlands}
\author{Eveline Klute}
\affiliation
{Department of Chemistry \& Pharmaceutical Sciences and Amsterdam Institute of Molecular and Life Sciences (AIMMS), Faculty of Science, Vrije Universiteit, De Boelelaan 1083, 1081HV Amsterdam, The Netherlands}
\author{Michael Seidl}
\affiliation
{Department of Chemistry \& Pharmaceutical Sciences and Amsterdam Institute of Molecular and Life Sciences (AIMMS), Faculty of Science, Vrije Universiteit, De Boelelaan 1083, 1081HV Amsterdam, The Netherlands}
\author{Paola Gori-Giorgi*}
\affiliation{Department of Chemistry \& Pharmaceutical Sciences and Amsterdam Institute of Molecular and Life Sciences (AIMMS), Faculty of Science, Vrije Universiteit, De Boelelaan 1083, 1081HV Amsterdam, The Netherlands}
\affiliation{Microsoft Research AI for Science, Evert van de Beekstraat 354, 1118CZ Schiphol, The Netherlands}
\affiliation{p.gorigiorgi@vu.nl}

\title{The M{\o}ller-Plesset Adiabatic Connection at Large Coupling Strengths for Open-shell Systems}

\begin{abstract}
\textbf{Abstract:} We study the adiabatic connection that has as weak-coupling expansion the M{\o}ller-Plesset perturbation series, generalizing to the open-shell case previous closed-shell results for the large-coupling limit. We first focus on the hydrogen atom with fractional spins, providing results along the adiabatic connection, from small to large coupling strengths. We reveal an intriguing phase diagram, and an equation for the large-coupling leading order that has closed-form solutions for specific choices of its relevant quantum numbers. We then show that the hydrogen atom results provide variational estimates for the large-coupling leading terms for the general many-electron open-shell case in terms of functionals of the Hartree-Fock $\alpha$-spin and $\beta$-spin densities.  
\end{abstract}

\maketitle

%%%%%%%%%%%%%%%%%%%%%%%%%%%%%%%%%%%%%%%%%%%%%%%%%%%%%%%%%%%%%%%%%%%%%
%% The manuscript does not need to include \maketitle, which is
%% executed automatically.  The document should begin with an
%% abstract, if appropriate.  If one is given and should not be, the
%% contents will be gobbled.
%%%%%%%%%%%%%%%%%%%%%%%%%%%%%%%%%%%%%%%%%%%%%%%%%%%%%%%%%%%%%%%%%%%%%
\section{Introduction}

Understanding and being able to compute the effects of spin within any approximate many-electron framework plays a crucial role in the development of new quantum chemical methods.\cite{JacRei-IJQC-12,Rei-CHI-09,CraTru-PCCP-09,Gho-JBIC-06,TsuScu-JCP-09,TsuScu-JCP-10,TsuHenScuSav-JCP-10,EllMarScu-JCTC-13,BulHenScu-JCTC-15,WenMarHenScu-CR-12,HenCheScu-JCP-22,CohMorYan-CR-12,MegaPaper-PCCP-22} Already in the simple hydrogen molecule, spin plays an important role, ranging from restricted Hartree Fock (RHF) having the wrong behaviour for the total energy in the dissociation limit, to unrestricted Hartree Fock (UHF) breaking the spin symmetry to fix it, but it also provides a paradigmatic case to understand static correlation errors of approximate Density Functional Theory (DFT) functionals.\cite{VydScuPer-JCP-07,CohMorYan-SCI-08,CohMorYan-JCP-08,CohMorYan-JCTC-09,CohMorYan-CR-12,Sav-CP-09,VucWagMirGor-JCTC-15,BurMarDaaGorLoo-JCP-21,MegaPaper-PCCP-22} The so-called flat plane conditions\cite{CohMorYan-SCI-08,CohMorYan-JCP-08,CohMorYan-JCTC-09,CohMorYan-CR-12,Sav-CP-09,BurMarDaaGorLoo-JCP-21,MegaPaper-PCCP-22} that can guide the construction of approximate functionals\cite{BajJanKul-JCP-17,YanPatQuiZadEspAye-JCP-16,SuLiYan-PNAS-18} were derived by using the prototypical example of the spin dependence in the H atom.

Recently, the strong-coupling limit of the adiabatic connection (AC) that links the Hartree-Fock system to the physical one with the Møller-Plesset (MP) pertubation series as its weak coupling limit, has been studied in detail, giving the exact result for the leading order term and variational estimates for the next two orders in the closed-shell case.\cite{Per-IJQC-18,SeiGiaVucFabGor-JCP-18,DaaGroVucMusKooSeiGieGor-JCP-20,BurMarDaaGorLoo-JCP-21,DaaKooGroSeiGor-JCTC-22,GiaPri-JCP-22,GiaPri-JCTC-23} All three of these terms are functionals of the HF density, $\rho^{\rm HF}$, only. Using these results, functionals that interpolate between the weak and strong coupling limits of the MPAC have been introduced, using the exact HF exchange and MP2 correlation energies, combined with the strong-coupling limit.\cite{DaaFabDelGorVuc-JPCL-21} These interpolated functionals have been shown to massively improve MP2 interaction energies for a wide variety of non-covalent interactions, ranging from small charge-transfer complexes to larger $\pi-\pi$ bonded complexes.\cite{DaaFabDelGorVuc-JPCL-21} Generalizations to include other variants of MP2, such as opposite-spin only and regularization, have been found to be even more accurate at a lower computational cost.\cite{DaaKooPetFabDelGorVuc-JPCL-23}  

All these interpolation ideas originated from the DFT AC where, instead, the Kohn-Sham system is connected to the physical system, and the strong coupling limit is given by the strictly correlated electrons (SCE) state.\cite{Sei-PRA-99,SeiGorSav-PRA-07,Lew-CRM-18,CotFriKlu-ARMA-18,GroKooGieSeiCohMorGor-JCTC-17,GroSeiGorGie-PRA-19,ColDimStra-arxiv-21,SeiPerLev-PRA-99,SeiPerKur-PRL-00,GorVigSei-JCTC-09,LiuBur-PRA-09,VucGorDelFab-JPCL-18,GiaGorDelFab-JCP-18,Con-PRB-19,GiaPri-JCTC-23} A fundamental difference between the two ACs is that in the DFT one the density remains fixed as the coupling constant $\lambda$ is turned on whereas in the MPAC the density can roam freely. In the DFT AC the role of the spin state has been found to enter in the large coupling limit only at orders\cite{GorSeiVig-PRL-09,GorVigSei-JCTC-09,GroKooGieSeiCohMorGor-JCTC-17} $\sim e^{-\sqrt{\lambda}}$, which means that it can be ignored in the two leading $\lambda\to\infty$ terms used in the interpolating functionals.

However, in the MPAC case the spin state already affects the second leading term at strong coupling, due to the role of the exchange operator at this order and the lack of the density constraint.\cite{DaaGroVucMusKooSeiGieGor-JCP-20,BurMarDaaGorLoo-JCP-21} This spin-dependence becomes easy to study in the closed-shell case, where it has been shown\cite{DaaGroVucMusKooSeiGieGor-JCP-20} that the result for the H atom with $\frac12$ spin-up and $\frac12$ spin-down electrons (denoted H$[\frac12,\frac12]$) provides a variational estimate for the general many-electron case.  However, this unnecessarily restricts the chemical space that the new MPAC functionals can be used for.  Studying how the spin affects the MPAC strong coupling limit, but also its role along the whole adiabatic connection path beyond the closed-shell case, is the gap that we fill in this work. 

The paper starts with an introduction of the MPAC in Sec.~\ref{sec:theory}, including a summary of previous results for its strong coupling limit. Since the closed-shell many-electron case was obtained by generalizing the result for the H$[\frac12,\frac12]$, we start by studying the H atom MPAC beyond the spin-unpolarized case in Sec.~\ref{sec:MPACHatom}, where we derive and solve numerically the relevant equations, revealing an interesting phase diagram along the AC path. The spin-dependence of the strong-coupling limit coefficients is then extracted in section~\ref{sec:HatomLargeLambda}. As we shall see, this limit defines an equation that has closed form solutions only for some special values of its parameters (orbital angular momentum and spin). We then show in Sec.~\ref{sec:manyelectrons} that the results for the H atom with fractional spins provide a variational estimate for the strong-coupling MPAC functionals for the general many-electron case, in terms of the HF $\alpha$-spin and $\beta$-spin densities. Conclusions and perspectives are discussed in Sec.~\ref{sec:conc}.

%However, this is too expensive or even impossible in practice for systems larger than the hydrogen-atom, so functionals that approximate $W_{c}$ or directly $E_c$ have been introduced. Although these functionals are shown to be competitive with standard DFT functionals for non-covalent interactions, they contain empirical parameters that are fitted on a training set. In order to decrease the number of parameters and to increase the accuracy further, more exact information can be used, e.g. using the GEA's for $W_{\infty}$ and $W_{12}$, but also a possible extension to the complex plane to find critical points.

%Another feature that could increase the applicability of the MPAC functionals is by studying the effect of spin, because the current functionals can only be used for closed shell systems. By understanding how the MPAC curves depend on spin can give us insight how these functionals should act for open shells systems. 

\section{M{\o}ller-Plesset adiabatic connection}\label{sec:theory}

For a system with $N_\alpha$ spin-up and $N_\beta$ spin-down electrons in a given external potential $v_{\rm ext}(\rv)$, a Hartree Fock (HF) calculation amounts to minimizing the expectation value of the physical Hamiltonian over single Slater determinants only, yielding $N=N_\alpha+N_\beta$ occupied HF spin orbitals $\phi_i^{\rm HF}(\rv,\sigma)=\phi_i^{\rm HF}(\xv)$, with the HF electron density and the $\alpha$-spin and $\beta$-spin densities
\begin{align}
\rho^{\rm HF}(\rv) & =\rho_\alpha^{\rm HF}(\rv)+\rho_\beta^{\rm HF}(\rv), \nonumber \\
\rho_\sigma^{\rm HF}(\rv) & =\sum_{i=1}^{N_\sigma}|\phi_i^{\rm HF}(\rv,\sigma)|^2.
\label{rhoHFgen}\end{align}
We keep the notation general, such that the following equations apply both to restricted (the spatial part of the $\alpha$ and $\beta$ orbitals is forced to stay the same) and unrestricted open-shell HF.

Fixed in terms of these spin-orbitals (which are determined in the initial HF calculation), the standard Hartree and exchange operators $\hat{J}=\hat{J}[\rho^{\rm HF}]$ and $\hat{K}=\hat{K}[\{\phi_i^{\rm HF}\}]$ (that appear in the initial HF equations) do not change along the adiabatic connection defined below. Subsequently, treating $\hat{J}$ and $\hat{K}$ as fixed ($\lambda$-independent) one-body operators, the M\o ller Plesset adiabatic connection (MPAC) for this $N$-electron system is represented by the generalized ($\lambda$-dependent) Hamiltonian
\begin{equation}
\hat{H}^{\rm HF}_{\lambda}
= \hat{T}+\hat{V}_{\rm ext}+\hat{J}-\hat{K} + \lambda\Big(\hat{V}_{\rm ee} - \hat{J}+\hat{K}\Big),
\label{HamHFlambdaN}\end{equation}
where $\hat{T}$, $\hat{V}_{\rm ext}$, and $\hat{V}_{\rm ee}$, respectively, are the  kinetic energy, external potential $\sum_i v_{\rm ext}(\rv_i)$\footnote{Although spin-dependent potentials could be included in the MPAC, we are excluding it since we are staying within the RHF framework.}, and two-body electron-electron repulsion operators. Notice that $\hat{H}^{\rm HF}_{\lambda}$ for $\lambda\ne0$ (and $N\ge2$) is no longer a true HF (one-body) Hamiltonian, but includes a two-body interaction $\lambda\hat{V}_{\rm ee}$. 

We denote the ground state of $\hat{H}^{\rm HF}_{\lambda}$ by $|\Psi_{\lambda}\rangle$ and its corresponding eigenvalue by $E^{\rm HF}_{\lambda}$,
\begin{equation}
\hat{H}^{\rm HF}_{\lambda}\,|\Psi_{\lambda}\rangle=E^{\rm HF}_{\lambda}\,|\Psi_{\lambda}\rangle.
\nonumber\end{equation}
The Hellmann-Feynman theorem implies
\begin{equation}
\frac{{\rm d}E^{\rm HF}_{\lambda}}{{\rm d}\lambda}
=\Big\langle\Psi_{\lambda}\Big|\hat{V}_{\rm ee}-\hat{J}+\hat{K}\Big|\Psi_{\lambda}\Big\rangle.
\label{WclamDef0}\end{equation}
While $\hat{H}^{\rm HF}_{\lambda=1}=\hat{T}+\hat{V}_{\rm ee}+\hat{V}_{\rm ext}$ is the physical Hamiltonian (including the two-body interaction $\hat{V}_{\rm ee}$), the one-body operator $\hat{H}^{\rm HF}_{\lambda=0}=\hat{T}+\hat{V}_{\rm ext}+\hat{J}-\hat{K}$ is the original HF Hamiltonian whose ground state $|\Psi_{0}\rangle=|\Psi_{\lambda=0}\rangle$ is the Slater determinant made of the occupied HF spin-orbitals $\{\phi^{\rm HF}_i\}$. The HF energy of the $N$-electron system is defined as $E^{\rm HF}=\langle\Psi_0|\hat{H}^{\rm HF}_{\lambda=1}|\Psi_0\rangle$,
\begin{equation}
E^{\rm HF}=E^{\rm HF}_{\lambda=0}-\Big(U[\rho^{\rm HF}]+E_{\rm x}[\{\phi_i^{\rm HF}\}]\Big),
\nonumber\end{equation}
where the Hartree energy $U[\rho^{\rm HF}]=\frac12\langle\Psi_0|\hat{J}|\Psi_0\rangle$ is an explicit density functional,
\begin{equation}
U[\rho]=\frac12\int\!\!\!\int{\rm d}\rv\,{\rm d}\rv'\,\frac{\rho(\rv)\,\rho(\rv')}{|\rv-\rv'|},
\label{hartreeUdef}\end{equation}
while $E_{\rm x}[\{\phi_i^{\rm HF}\}]=-\frac12\langle\Psi_0|\hat{K}|\Psi_0\rangle<0$ is the usual HF exchange energy. 
The difference between the physical ground-state energy $E_{\rm exact}=E^{\rm HF}_{\lambda=1}$ and $E^{\rm HF}$ is the HF correlation energy,
\begin{equation}\label{eq:HellFeyn}
E^{\rm HF}_{\rm c}=E_{\rm exact}-E^{\rm HF}=\int_0^1{\rm d}\lambda\,W_{{\rm c},\lambda}.
\nonumber\end{equation}
Here we have introduced the MPAC integrand %(see Fig.~2 for illustration)
\begin{equation}
W_{{\rm c},\lambda}=\frac{{\rm d}E^{\rm HF}_{\lambda}}{{\rm d}\lambda}
+\Big(U[\rho^{\rm HF}]+E^{\rm HF}_{\rm x}[\{\phi_i^{\rm HF}\}]\Big)
\label{WclamDef}\end{equation}
where we have Eq.~\eqref{WclamDef0} for $\frac{{\rm d}E^{\rm HF}_{\lambda}}{{\rm d}\lambda}$.
The Taylor expansion of $W_{{\rm c},\lambda}$ for $\lambda\to0$ is the MP perturbation series
\begin{equation}\label{eq:WHFMP}
W_{c,\lambda\rightarrow 0}=\sum_{n=2}^\infty n\,E^{{\rm MP}n}_{c}\,\lambda^{n-1}.
\end{equation}
This expansion holds for closed systems. The first part of this work, however, addresses open fragments of larger systems (for example an H atom within an infinitely stretched H$_2$ molecule). In such cases, Eq.~\eqref{eq:WHFMP} applies to the whole system while for the subsystem we may find $W_{c,\lambda=0}\ne0$. An explicit example is reported in Fig. 10 of Ref.~\onlinecite{DaaGroVucMusKooSeiGieGor-JCP-20}, where it is shown that the MPAC result for the stretched H$_2$ molecule tends to twice the result for the H atom with $\frac12$ spin-up and $\frac12$ spin-down as the distance $R$ between the two H atoms increases, except at $\lambda=0$, where the order of limits, $R\to\infty$ and $\lambda\to 0$, matters.

\subsection{Previous results on the $\lambda\to\infty$ expansion}\label{sec:A} % A
The counterpart of Eq.~\eqref{eq:WHFMP} is the large coupling strength ($\lambda\to\infty$) expansion \cite{SeiGiaVucFabGor-JCP-18,DaaGroVucMusKooSeiGieGor-JCP-20}
\begin{align}\label{eq:finalexp}
    W_{c,\lambda\rightarrow\infty} & = W_{c,\infty} + \frac{W_{\frac{1}{2}}}{\sqrt{\lambda}}+\frac{W_{\frac{3}{4}}}{\lambda^{\frac{3}{4}}}+\dots 
\end{align}
For general $N$-electron systems (atoms or molecules with $M$ fixed nuclear positions $\Rv_k$ with charge numbers $Z_k$, where $k=1,...,M$), the leading term is \cite{SeiGiaVucFabGor-JCP-18}
\begin{equation}
	W_{c,\infty} = E_{\rm el}[\rho^{\rm HF}]+E_{\rm x}[\{\phi_i^{\rm HF}\}].
\label{eq:Wcinffinal}\end{equation}
Here, $E_{\rm el}[\rho]$ is the classical electrostatic energy of $N$ negative point charges (classical electrons) sitting at equilibrium positions in a rigid continuous positive background charge distribution with given density $\rho(\rv)$,
\begin{equation}\label{eq:EelDef}
	E_{\rm el}[\rho]= \min_{\{\rv_1,...,\rv_N\}}\left\{\sum_{i>j=1}^{N}\frac{1}{|\rv_i-\rv_j|}-\sum_{i=1}^N v^{\rm H}\big(\rv_i;[\rho]\big)+U[\rho]\right\},
\end{equation}
with the electrostatic (Hartree) potential due to the charge distribution $\rho(\rv)$,
\begin{equation}
v^{\rm H}\big(\rv;[\rho]\big)=\int \frac{\rho(\rv')}{|\rv'-\rv|} {\rm d}\rv'.
\nonumber\end{equation}
While the minimizing set $\{\rv^{\rm min}_1,...,\rv^{\rm min}_N\}$ of equilibrium positions in Eq.~\eqref{eq:EelDef} is typically not unique (depending on the symmetry group of the molecule), 
the set of $N$ density values $\rho^{\rm HF}(\rv_i^{\rm min})$, for $i=1,...,N$, should be unique.
Independently, we expect a certain number $I_{\rm nuc}\le N$ of these positions $\rv^{\rm min}_i$ to coincide with some of the fixed nuclear positions $\Rv_k$ (with $k=1,...,M$), implying that $I_{\rm nuc}\le M$. Then, after re-labeling the $\rv^{\rm min}_i$ if necessary, we have  
\begin{equation}
\rv^{\rm min}_i=\Rv_{k_i}\qquad(i=1,...,I_{\rm nuc}).
\label{eq:min14}\end{equation}
In terms of these values $\rho^{\rm HF}(\rv_i^{\rm min})$, the coefficients of the remaining terms in Eq.~\eqref{eq:finalexp} have been shown\cite{DaaGroVucMusKooSeiGieGor-JCP-20} to have the variational estimate for {\em closed-shell} $N$-electron systems
\begin{align}
W_{\frac12} & \approx \frac{\tilde{\epsilon}_{\frac12}}{2} \sum_{i=1}^N\sqrt{4\pi\rho^{\rm HF}(\rv_i^{\rm min})} \label{eq:Wc1/2fin}\\
W_{\frac34} & \approx \frac{\tilde{\epsilon}_{\frac14}}{4}\sum_{i=1}^{I_{\rm nuc}}Z_{k_i}  \sqrt[4]{4\pi\rho^{\rm HF}\big(\Rv_{k_i}\big)}
\label{eq:finalambda3/4}
\end{align}
Here, $\tilde{\epsilon}_{\frac{1}{2}}=1.6185$ and $\tilde{\epsilon}_{\frac{1}{4}}=-2.70306$ are the values labeled ``$s=\frac12$'' in {\color{blue} Tab.~\ref{tab:tabs1s12}}. 

Eqs.~\eqref{eq:Wc1/2fin} and \eqref{eq:finalambda3/4} for closed-shell systems were obtained in Ref.~\onlinecite{DaaGroVucMusKooSeiGieGor-JCP-20} by generalizing the exact coefficients for a closed-shell version ($s=\frac12$) of the hydrogen atom,
%($N=1$, $\hat{V}_{\rm ee}=0$, $\hat{V}_{\rm ext}=-\frac{Z}r$) represented by an ensemble of two states in the spin-unpolarized case ($s=\frac12$),
\begin{align}
	W_{c,\infty}(s)    & = -v^{\rm H}\big({\bf 0};[\rho_s^{\rm HF}]\big)+(1-s)\,U[\rho_s^{\rm HF}] \label{eq:WcinffinalN1}\\
	W_{\frac{1}{2}}(s) & =  \frac{\tilde{\epsilon}_{\frac12}(s)}2\sqrt{4\pi\rho_s^{\rm HF}({\bf 0})} \label{eq:Wc1/2finN1}\\
	W_{\frac{3}{4}}(s) & = Z\frac{\tilde{\epsilon}_{\frac14}(s)}4\sqrt[4]{4\pi\rho_s^{\rm HF}({\bf 0})}\label{eq:finalambda3/4N1}
\end{align}
For the hydrogen atom ($N=1$, $\hat{V}_{\rm ee}=0$, $\hat{V}_{\rm ext}=-\frac{Z}r$), Eq.~\eqref{eq:EelDef} has $\sum_{i>j=1}^{N}\frac{1}{|\rv_i-\rv_j|}=0$, and the only minimizing position $\rv_1^{\rm min}={\bf 0}$ coincides with the only nuclear position $\Rv_1={\bf 0}$, where minus the Hartree potential has its minimum. In Ref.~\onlinecite{DaaGroVucMusKooSeiGieGor-JCP-20}, only the two cases $s=1$ of a spin-polarized regular atom H$[1,0]$ and $s=\frac12$ of a fully spin-unpolarized ensemble H$[\frac12,\frac12]$  were studied.
The values of the three quantities $\rho^{\rm HF}(\rv)$, $\tilde{\epsilon}_{\frac12}$, and $\tilde{\epsilon}_{\frac14}$ are $s$ dependent, with the latter two for $s\in\{\frac12,1\}$ reported again in Table~\ref{tab:tabs1s12} for completeness. 

\begin{table}[t]
    \centering
\begin{tabular}{|c||r|r|}\hline
                                 & $s=\frac12$ & $s=1$ \\\hline\hline
$\tilde{\epsilon}_{\frac{1}{2}}(s)$ & 1.6185\hspace*{1.9mm} & 2.0207 \\\hline
$\tilde{\epsilon}_{\frac{1}{4}}(s)$ & $-$2.70306 & $-$3.2009 \\\hline
\end{tabular}
    \caption{The value of $\tilde{\epsilon}_{\frac{1}{2}}(s)$ and $\tilde{\epsilon}_{\frac{1}{4}}(s)$ for the hydrogen atoms with weight factor $s=1$ and $s=\frac{1}{2}$ from Ref.~\onlinecite{DaaGroVucMusKooSeiGieGor-JCP-20}.}
    \label{tab:tabs1s12}
\end{table}

As a particular feature, the expansion \eqref{eq:finalexp} has a term $O(\lambda^{-\frac34})$. Such a term is absent in the corresponding $\lambda\to\infty$ expansion for the density-fixed adiabatic connection in DFT. According to Eq.~\eqref{eq:finalambda3/4}, the term $O(\lambda^{-\frac34})$ in Eq.~\eqref{eq:finalexp} occurs only in molecules with $I_{\rm nuc}>0$. For a short explanation, \cite{DaaGroVucMusKooSeiGieGor-JCP-20} we note that the spatial probability distribution of the $N$ electrons in the state $|\Psi_{\lambda}\rangle$ for $\lambda\to\infty$ concentrates around the positions $\rv^{\rm min}_i$. Consequently, each singularity $-\frac{Z_{k_i}}{|\rv-\Rv_{k_i}|}$ (with $1\le i\le I_{\rm nuc}$) of the external potential $\hat{V}_{\rm ext}$ in $\hat{H}^{\rm HF}_{\lambda}$ of Eq.~\eqref{HamHFlambdaN}
contributes $O(\lambda^{\frac14})$ to $E^{\rm HF}_{\lambda}$ as $\lambda\to\infty$ and therefore $O(\lambda^{-\frac34})$ to $W_{c,\lambda}$. Moreover, due to the Kato cusps of $\rho^{\rm HF}(\rv)$ at nuclear positions, the term $\lambda(-\hat{J}+\hat{K})$ in $\hat{H}^{\rm HF}_{\lambda}$ for each $i\le I_{\rm nuc}$ produces two additional contributions $O(\lambda^{\frac14})$ to $E^{\rm HF}_{\lambda}$. The result for the H$[\frac12,\frac12]$ case, was proven to yield variational estimates for the many-electron closed-shell case.\cite{DaaGroVucMusKooSeiGieGor-JCP-20}

\section{M{\o}ller-Plesset adiabatic connection for the H atom with fractional spin}
\label{sec:MPACHatom}
In this section we generalize and compute the MPAC for the hydrogen atom at arbitrary values $\frac12\le s\le1$ of the weight parameter $s$. This parameter $s$, defined in Eq.~\eqref{swDef} below, must not be confused with the spin quantum number. In Sec.~\ref{sec:manyelectrons}, we show that in the large-$\lambda$ limit of the MPAC for a general open-shell system the parameter $s$ is linked to the local spin polarization, Eq.~\eqref{eq:szeta}.

Here we obtain results of the MPAC for $\lambda\in[0,\infty)$ for the H atom with general $s$, which, as we shall see, reveal an interesting phase diagram.

\subsection{The HF orbital $\phi_s(\rv)$}\label{sec:B}  % B
We consider a  one-electron system ($\hat{V}_{\rm ee}=0$) in a hydrogen type external potential $
v_{\rm ext}(\rv)=-\frac{Z}r$.
Instead of being in a pure quantum state, however, this system is described by an ensemble of a spin-up state 
$\phi_\alpha(\rv)|\alpha\rangle$ and a spin-down state $\phi_\beta(\rv)|\beta\rangle$ with weights $1-w$ and $w$, respectively. Statistically, our system has spin
\begin{equation}
m_{\rm s}\;=\;\textstyle\frac12-w\qquad\qquad(0\le w\le1).
\label{msFrac}\end{equation}
In this study we stay in a restricted open-shell HF (ROHF) framework, forcing both spin states to have the same \textit{real} spatial orbital $\phi(\rv)=\phi_{s}(\rv)$, which will depend on the weight $w$ via\cite{BurMarDaaGorLoo-JCP-21} the parameter $s$, 
\begin{equation}
%s_w=1-2w+2w^2\qquad\big({\textstyle\frac12}\le s_w\le1\big).
s=1-2w\,(1-w)\qquad\big({\textstyle\frac12}\le s\le1\big),
\label{swDef}\end{equation}
and is fixed by minimizing the weight-dependent functional \cite{BurMarDaaGorLoo-JCP-21}
\begin{equation}
{\cal E}_{s}^{\rm ROHF}[\phi]\;=\;\big\langle\phi\big|\hat{T}+\hat{V}_{\rm ext}\big|\phi\big\rangle
\,+\,U[\phi^2]\,+\,E_{{\rm x},s}[\phi].
\nonumber\end{equation}
The influence of the restricted open-shell choice made here for the generalization to the many electron open shell case to the unrestricted case is discussed in Sec.~\ref{sec:manyelectrons}.

Then, $\phi_{s}(\rv)^2=\rho_{s}^{\rm HF}(\rv)$ will be the HF density of Eq.~\eqref{rhoHFgen}, with the Hartree energy $U[\rho_{s}^{\rm HF}]$. The exchange functional for this ensemble system is explicitly weight-dependent \cite{BurMarDaaGorLoo-JCP-21}
\begin{equation}
E_{{\rm x},s}[\phi] = -s\,U[\phi^2].
\nonumber\end{equation}
In the pure-state cases ($w=0$ or $w=1$)  $E_{{\rm x},s}[\phi]$ exactly compensates the spurious Hartree interaction $U[\phi^2]$. In the ensemble case ($0<w<1$) this compensation is incomplete. \cite{CohMorYan-SCI-08,BurMarDaaGorLoo-JCP-21}
This one-electron RHF functional of $\phi$ explicitly reads 
\begin{equation}
{\cal E}_s^{\rm RHF}[\phi] = \big\langle\phi\big|\hat{T}+\hat{V}_{\rm ext}\big|\phi\big\rangle
\,+\,\frac{1-s}2\int\!\!\!\int{\rm d}\rv\,{\rm d}\rv'\,\frac{\phi(\rv)^2\,\phi(\rv')^2}{|\rv-\rv'|}.
\label{EphiFix}\end{equation}
In cases with $s\ne1$, the spherically symmetric (real-valued) minimizer
\begin{align}
\phi_s(\rv) & = {\rm arg}\min_{\phi(\rv)}\,{\cal E}_s^{\rm RHF}[\phi]\nonumber\\
& = R_s(r)\,Y_{00}(\theta,\varphi)\equiv\frac{R_s(r)}{\sqrt{4\pi}}
\label{phiFix}\end{align}
will be {\em different} from the hydrogen ground state $\psi_{\rm 1s}(\rv)$. However, the radial wave function $R_s(r)$ is still finite at $r=0$ and satisfies Kato's cusp condition $R'_s(0)=-ZR_s(0)$.\footnote{Notice that Eq.~(20) in Ref.~20 uses for $R_s(r)$ the notation $\phi_s(r)=\sqrt{4\pi}\phi_s(\rv)$. To avoid confusion between the functions $\phi_s(r)$ and $\phi_s(\rv)$, we have modified here our notation.}

The non-linear SCF (self consistent field) Euler-Lagrange equation for $\phi_s(\rv)$ corresponding to the minimization in Eq.~\eqref{phiFix},
\begin{equation}
-\frac12\nabla^2\phi(\rv)\,-\,\frac{Z}r\phi(\rv)\,+\,(1-s)\int{\rm d}\rv'\,\frac{\phi(\rv')^2}{|\rv-\rv'|}\,\phi(\rv)=\epsilon\,\phi(\rv),
\nonumber\end{equation}
is solved using a basis set expansion in terms of Slater type orbitals (STOs) as in Ref.~\onlinecite{DaaGroVucMusKooSeiGieGor-JCP-20},
\begin{equation}
R_s(r)=\sum_{n=1}^{10}c_nr^{n-1}{\rm e}^{-r}.
\label{ansatzSTO}\end{equation}

\subsection{M{\o}ller-Plesset adiabatic connection: equations and numerical solutions}
Employing the HF orbital $\phi_s(\rv)$, fixed by Eq.~\eqref{phiFix} for a given value of $s\in [\frac12,1]$, we now consider the $\lambda$-dependent Hamiltonian of Eq.~\eqref{HamHFlambdaN} for one-electron systems ($\hat{V}_{\rm ee}=0$),
\begin{equation}
\hat{H}^{\rm HF}_{s,\lambda}=\hat{T} + \hat{V}_{\rm ext} + (1 - \lambda)\Big(\hat{J}[\phi_s] - \hat{K}_s[\phi_s]\Big)\qquad(\lambda\ge0).
\label{HamHFlambda}\end{equation}

\subsubsection{The operators $\hat{J}$ and $\hat{K}_s$} % C.1

For the present case of one-electron systems with fractional spin, the Hartree operator $\hat{J}[\phi]$ and the (explicitly weight-dependent) exchange operator $\hat{K}_s[\phi]$ are defined by their action,
\begin{align}\label{JKwDef}
\hat{J}[\phi]\,\Psi(\xv) & = \left[\int{\rm d}\rv'\,\frac{|\phi(\rv')|^2}{|\rv-\rv'|}\right]\,\Psi(\xv)\nonumber\\
\hat{K}_{s}[\phi]\,\Psi(\xv) & = \phi(\rv)\bigg[(1-w)\langle\sigma|\alpha\rangle\int{\rm d}\rv'\,\frac{\phi^*(\rv')\psi^\alpha(\rv')}{|\rv-\rv'|}\nonumber\\
& +\,w\langle\sigma|\beta\rangle\int{\rm d}\rv'\,\frac{\phi^*(\rv')\psi^\beta(\rv')}{|\rv-\rv'|}\bigg],
\end{align}
on a general (single-particle) spin orbital
%\begin{equation}
%\Psi(\xv) \;=\; \psi^\alpha(\rv)\,|\alpha\rangle\,+\,\psi^\beta(\rv)\,|\beta\rangle 
%\;=\; \left(\begin{array}{c}\psi^\alpha(\rv)\\\psi^\beta(\rv)\end{array}\right).
%\nonumber\end{equation}

$$
    \Psi(\mathbf{x}) = \psi^{\alpha}(\mathbf{r})\langle\sigma\vert\alpha\rangle + \psi^{\beta}(\mathbf{r})\langle\sigma\vert\beta\rangle.$$
%with $\sigma$ being the spin of the minimizing wavefunction.
Notice that $\hat{J}[\phi]=\hat{J}[\phi_s]$ is purely multiplicative, $\hat{J}[\phi_s]\Psi(\xv)=v^{\rm H}_s(\rv)\Psi(\xv)$, with the weight-dependent Hartree potential $v^{\rm H}_s(\rv)=v^{\rm H}\big([\phi_s^2];\rv\big)$
\begin{equation}
v^{\rm H}_s(\rv)=\int{\rm d}\rv'\,\frac{\phi_s(\rv')^2}{|\rv-\rv'|},
\label{vHartree}\end{equation}
while $\hat{K}_s[\phi]$ is non-local and, in addition, acts on the spin variable $\sigma$ in $\Psi(\xv)=\Psi(\rv,\sigma)$.
Equivalently, writing $\hat{K}_s[\phi]\,\Psi(\xv)=\int{\rm d}\xv'\,k_s\big([\phi];\xv,\xv'\big)\Psi(\xv')$, the operator $\hat{K}_s[\phi]$ can be defined via its kernel
\begin{align}\label{eq:kernelK}
k_{s}\big([\phi];\textbf{x},\textbf{x}'\big) = & \frac{\phi(\rv)\phi^*(\rv')}{|\rv-\rv'|}\Big[(1-w)\,\braket{\sigma}{\alpha}\braket{\alpha}{\sigma’} \\
& +\,w\,\braket{\sigma}{\beta}\braket{\beta}{\sigma’} \Big].
\nonumber\end{align}

\subsubsection{Wave function for arbitrary $\lambda>0$} % C.2

In the case $\lambda=1$, $\hat{H}^{\rm HF}_{s,\lambda=1}=\hat{T}+\hat{V}_{\rm ext}$ is the physical Hamiltonian of the H atom, which has a spin-degenerate ground state. Any superposition of the kind
\begin{equation}
\Psi_{\lambda=1}(\xv)\;=\;\psi_{\rm 1s}(\rv)\Big(\sqrt{1-q}\,\langle\sigma\vert\alpha\rangle\,+\,\sqrt{q}\,\langle\sigma\vert\beta\rangle\Big)
\label{GS1}\end{equation}
where $\psi_{\rm 1s}(\rv)$ is the hydrogenic 1s orbital with $q\in [0,1]$ is a valid ground state. Alternatively, instead of the superposition we can consider again a statistical ensemble. 

Generalizing Eq.~\eqref{GS1} to cases with arbitrary $\lambda>0$, we write the ground state $\Psi_{s,\lambda}(\xv)$ of $\hat{H}^{\rm HF}_{s,\lambda}$ as
\begin{equation}
\Psi(\xv)\;=\;\psi(\rv)\Big(\sqrt{1-w}\,\langle\sigma\vert\alpha\rangle\,+\,\sqrt{w}\,\langle\sigma\vert\beta\rangle\Big),
\label{ansatzPsi}\end{equation}
where we have chosen $q=w$, forcing the spin expectation $\langle\hat{S}_z\rangle$ in this pure state to equal the ensemble average $m_{\rm s}=\frac12-w$. In other words: we suppress spin flip and we stay on the restricted open-shell curve by enforcing the spatial orbital to be the same for both spins.   We also notice that we could alternatively use an ensemble in Eq.~\eqref{ansatzPsi} instead of a superposition. This does not change the result for the energy along the adiabatic connection, since the exchange kernel of Eq.~\eqref{eq:kernelK}, being diagonal in the spin part, yields the same expectation value for a superposition or an ensemble. All the equations reported below, in which the spin-dependence is explicitly transformed into a weight dependence in the MPAC Hamiltonian, are thus the same whether for the wavefunction at $\lambda>0$ we use a superposition or an ensemble. The only constraint that matters is forbidding spin flip, which we enforce to keep the AC curve smooth. In fact, if we allow the spin to relax, the MPAC has a discontinuity\cite{DaaGroVucMusKooSeiGieGor-JCP-20,BurMarDaaGorLoo-JCP-21} as we cross $\lambda=1$ (except in the case $s=\frac{1}{2}$). Since our aim is to build interpolations by using the information at large $\lambda$, we want to follow the AC that connects smoothly the $\lambda\in[0,1]$ region with the $\lambda\to\infty$ limit.

With Eq.~\eqref{ansatzPsi}, the expectation of the Hamiltonian \eqref{HamHFlambda} can be written as
\begin{align}
{\cal E}_{s,\lambda}[\psi] & = \big\langle\Psi\big|\,\hat{H}^{\rm HF}_{s,\lambda}\,\big|\Psi\big\rangle\nonumber\\
& = \Big\langle\psi\,\Big|\,\hat{T}+\hat{V}_{\rm ext}\,+\,(1-\lambda)\Big(\hat{J}[\phi_s]-s\hat{K}[\phi_s]\Big)\,\Big|\,\psi\Big\rangle.
\label{expectationHFlambda}
\end{align}
Here, we have performed the spin summation, turning $\hat{K}_s[\phi]$ into the simpler operator $\hat{K}[\phi]$, which no longer acts on spin,
\begin{equation}
\hat{K}[\phi]\,\psi(\rv) = \phi(\rv)\int{\rm d}\rv'\,\frac{\phi^*(\rv')\psi(\rv')}{|\rv-\rv'|},
\label{K0}\end{equation}
but instead needs the weight parameter $s$ from Eq.~\eqref{swDef} as a prefactor in Eq.~\eqref{expectationHFlambda}. As mentioned before, this holds regardless of the choice of using for $\Psi(\xv)$ a superposition or an ensemble.

\subsubsection{Evaluation of the MPAC integrand $W_{{\rm c},\lambda}$} % C.3

To evaluate the weight-dependent MPAC integrand of Eq.~\eqref{WclamDef} for the present system,
\begin{equation}
W_{c,\lambda}=\frac{{\rm d}E^{\rm HF}_{s,\lambda}}{{\rm d}\lambda}
+\Big(U[\phi_s^2]+E_{{\rm x},s}[\phi_s]\Big),
\label{WclamHatom}\end{equation}
we need the ground state energy $E^{\rm HF}_{s,\lambda}$ of $\hat{H}^{\rm HF}_{s,\lambda}$,
\begin{equation}
E^{\rm HF}_{s,\lambda}\;=\;\min_{\psi(\rv)}{\cal E}_{s,\lambda}[\psi].
\label{minCond3A}\end{equation}
Since our (fixed) HF orbital $\phi_s(\rv)=R_s(r)Y_{00}(\theta,\varphi)=\frac{R_s(r)}{\sqrt{4\pi}}$ is spherically symmetric, the same is true for the Hartree potential in Eq.~\eqref{vHartree}, $v^{\rm H}_s(\rv)=v^{\rm H}_s(r)$, and the problem becomes block-diagonal in the orbital angular momentum  $\ell$. Therefore, the minimization \eqref{minCond3A}, is performed separately for each $\ell$, with wave function $\psi(\rv)=\frac{u_{\ell}(r)}r\,Y_{\ell 0}(\theta,\varphi)$ where $u_{\ell}(r)$ is the minimizer $u(r)$, with $\int_0^\infty{\rm d} r\,u(r)^2=1$ and $u(0)=0$, in
%Since our (fixed) HF orbital $\phi_s(\rv)=R_s(r)Y_{00}(\theta,\varphi)=\frac{R_s(r)}{\sqrt{4\pi}}$ is spherically symmetric, implying the same for the Hartree potential in Eq.~\eqref{vHartree}, $v^{\rm H}_s(\rv)=v^{\rm H}_s(r)$, the minimization \eqref{minCond3A}, in a first step, can be performed separately in different angular momentum channels $\ell$, each with wave function $\psi(\rv)=\frac{u_{\ell}(r)}r\,Y_{\ell 0}(\theta,\varphi)$ where $u_{\ell}(r)$ is the minimizer $u(r)$ in
\begin{align}
E_{s,\lambda}(\ell) = & \min_{u(r)}\int_0^\infty{\rm d}r\bigg[\frac{u'(r)^2}2
+\bigg(\frac{\ell(\ell+1)}{2r^2}-\frac{Z}r\nonumber\\
& +(1-\lambda)\,v^{\rm H}_s(r)\bigg)u(r)^2\hspace*{3mm}\nonumber\\
& -\,(1-\lambda)\frac{2s}{2\ell+1}\frac{R_s(r)}{r^\ell}\,u(r)\nonumber\\
& \qquad \times \int_0^r{\rm d}r'\,(r')^{\ell+1}R_s(r')\,u(r')\bigg].
\label{minmin}\end{align}
Then, the minimum \eqref{minCond3A} is obtained as
\begin{equation}
E^{\rm HF}_{s,\lambda}\;=\;\min_{\ell}E_{s,\lambda}(\ell).
\label{minCond3B}\end{equation}
For any fixed value of $s$, the $\lambda$-dependent minimizer $\ell=\ell_s(\lambda)$ can jump between different integers $\ell$, as $\lambda$ continuously grows from $\lambda=0$ to $\lambda=\infty$. Plotted versus $\lambda$, $E^{\rm HF}_{s,\lambda}$ will be continuous with possible kinks (and corresponding jumps in the derivative $\frac{{\rm d}E_{s,\lambda}}{{\rm d}\lambda}$).

The Euler Lagrange equation for the minimization \eqref{minmin}, \begin{widetext}
\begin{align}
E_{s,\lambda}(\ell)\,u(r) = -\frac{u''(r)}2+\left(\frac{\ell(\ell+1)}{2r^2}-\frac{Z}r+(1-\lambda)\,v^{\rm H}_s(r)\right)\,u(r)\nonumber\\
-s\,\frac{1-\lambda}{2\ell+1}R_s(r)\left[\frac1{r^\ell}\int_0^r{\rm d}r'(r')^{\ell+1}R_s(r')u(r')
+r^{\ell+1}\int_r^{\infty}\frac{{\rm d}r'}{(r')^{\ell}}R_s(r')u(r')\right],
\label{EulerLagrange-lambda}\end{align}
\end{widetext}
is solved for a given $\{s,\,\ell\}$ pair  with the spectral renormalization method,\cite{AblMus-OL-05,AblMus-PRL-13,AblMus-NL-16,GroMusSeiGor-JPCM-20} following the algorithm described in Ref.~\onlinecite{DaaGroVucMusKooSeiGieGor-JCP-20}. 

%In the supplementary information the python code is provided, which can also be used to calculate the crossings of state and both $\Tilde{\epsilon}_{\frac{1}{2}}(s)$ and $\Tilde{\epsilon}_{\frac{1}{4}}(s)$. This code uses the python implementation of the originally in MATLAB implemented functions called lagroots, poldif and lagroots~\cite{WeiRed-ATMS-00}.
\subsection{Results: spin-dependence along the MPAC}

\begin{figure*}[t]
\begin{subfigure}{.49\textwidth}
    \centering
    \includegraphics[width=.95\linewidth]{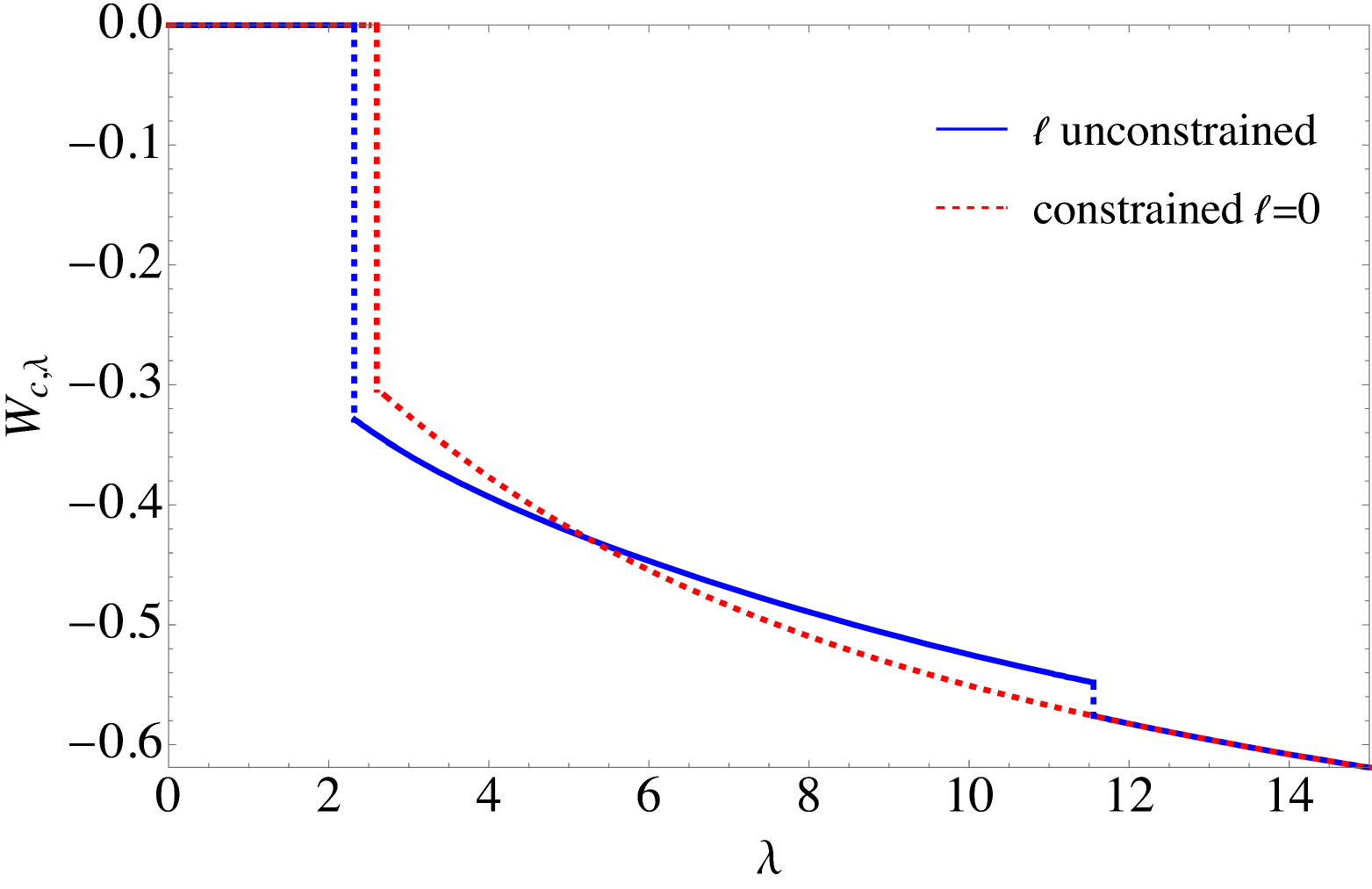}
\end{subfigure}
\begin{subfigure}{.49\textwidth}
    \centering
    \includegraphics[width=.95\linewidth]{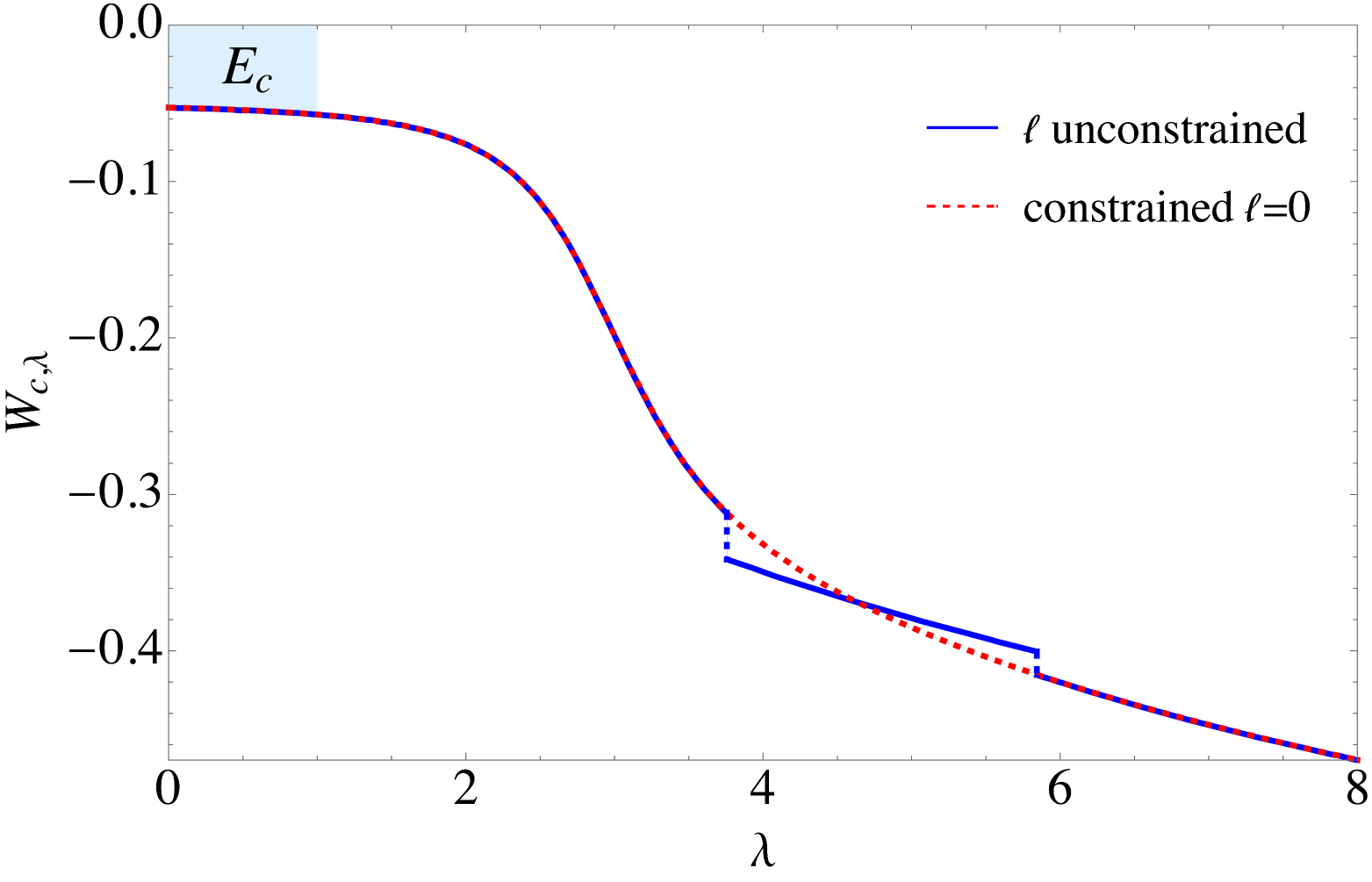}
\end{subfigure}
\begin{subfigure}{.49\textwidth}
    \centering
    \includegraphics[width=.95\linewidth]{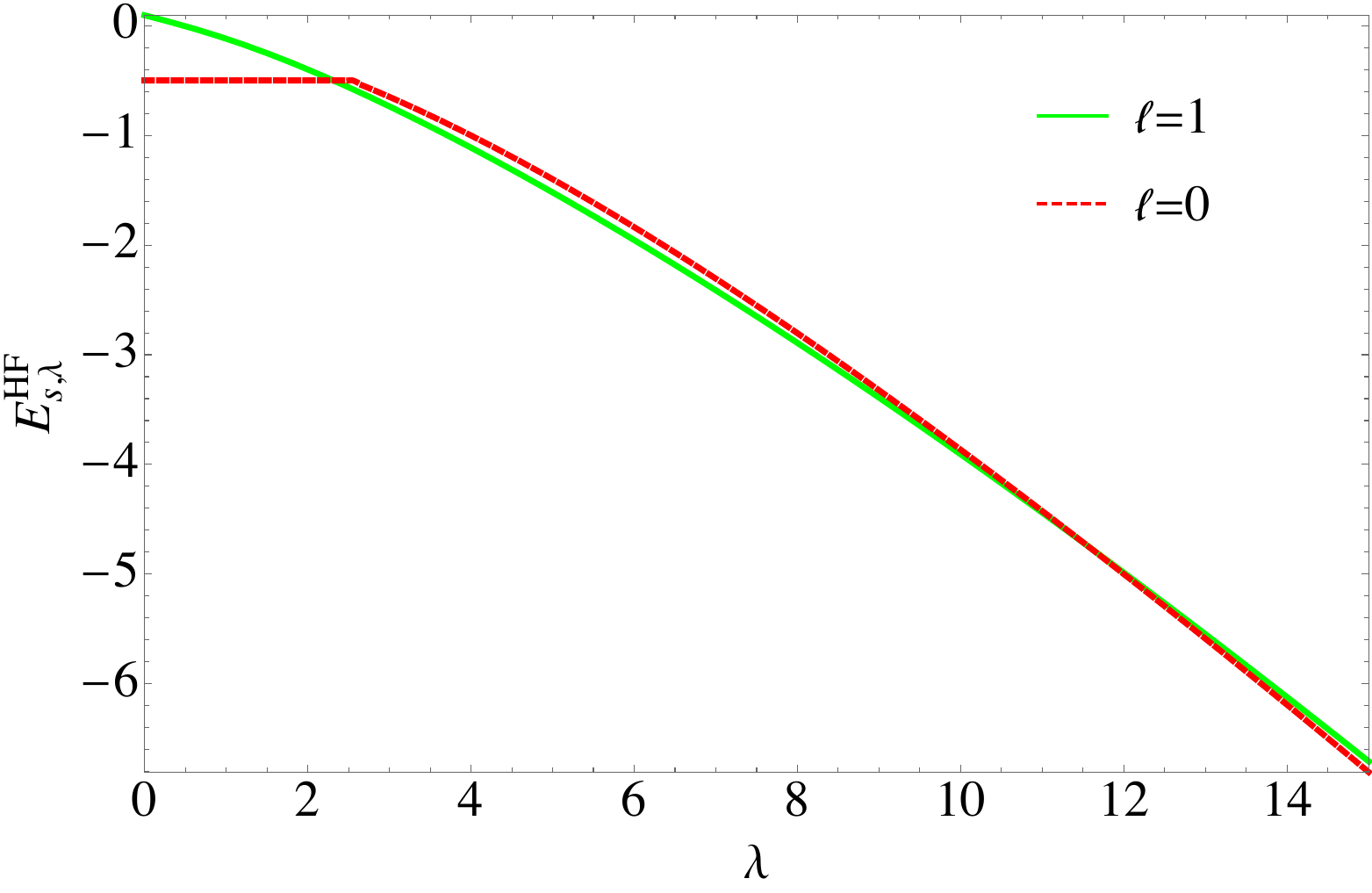}
\end{subfigure}
\begin{subfigure}{.49\textwidth}
    \centering
    \includegraphics[width=.95\linewidth]{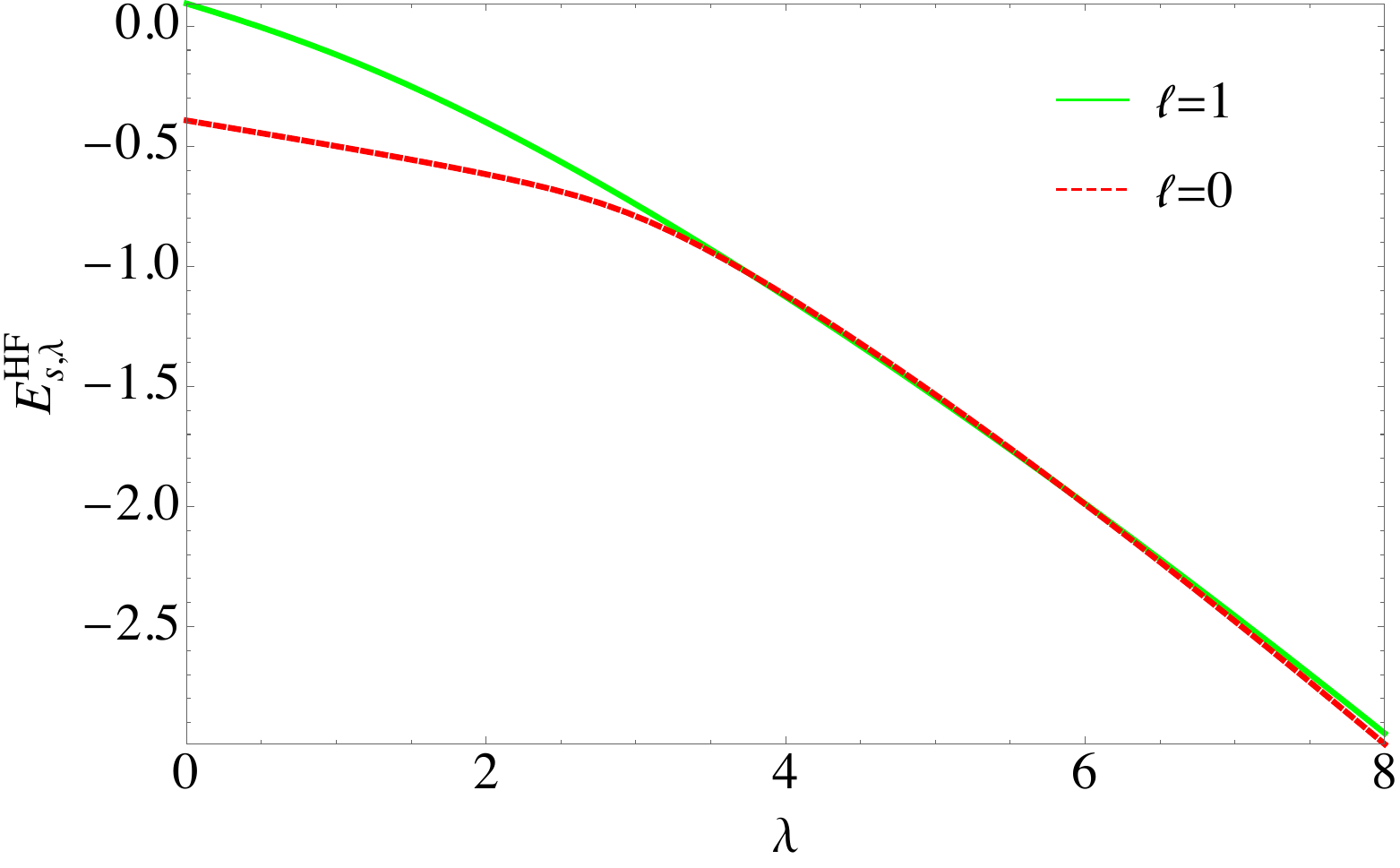}
\end{subfigure}
\caption{The $W_{c,\lambda}$ graph of $s=1$ ($w=\{0,1\}$) (top left) and $s=0.82$ ($w=\{0.1,0.9\}$) (top right) when $\ell$ is unconstrained and when we constrain $\ell=0$, with the blue shaded area being the correlation energy. The bottom two panels contain the corresponding $\ell=0$ and $\ell=1$ of $E_{s\lambda}^{\rm HF}$ curves showing the crossings of states between the two channels.}
\label{fig:EWgraphs}
\end{figure*}

In Ref.~\onlinecite{DaaGroVucMusKooSeiGieGor-JCP-20} it was found that for the spin-polarized ($s=1$, or $w=\{0,1\}$) H atom the lowest energy $E^{\rm HF}_{s=1,\lambda}$  starts at $\ell=0$ for small $\lambda$, then a first crossing of states from $\ell=0$ to $\ell=1$ occurs at $\lambda=2.3$, followed by a second crossing back to $\ell=0$ around $\lambda=11.5$, as shown here again on the bottom left panel of Fig.~\ref{fig:EWgraphs}. Instead, for the case $s=\frac12$, the $\ell=0$ state was found to be the lowest for all $\lambda\ge 0$. This means that the integrand $W_{c,\lambda}$ has discontinuities in the case $s=1$ (see, the top left panel Fig.~\ref{fig:EWgraphs}), while it is continuous for $s=\frac12$. 

The first question we address is thus whether the two crossings of states persist as we lower $s$ from 1 to $\frac12$ and how they eventually disappear at $s=\frac12$. We find that, as we lower $s$ starting from $s=1$, the region of $\lambda$ values for which $\ell=1$ is the ground state shrinks, until it disappears entirely at $s=0.810$. %In Fig.~\ref{fig:phase} we report the resulting phase diagram in the $\lambda,s$ plane, showing the region in which each $\ell$ gives the lowest energy. Values of $\ell$ larger than 1 are always found to be much higher in energy. Interestingly, the position of the second crossing is much more sensitive to changes in $s$ compared to the first crossing. This makes sense because the first crossing is a sharp crossing, which causes a large step in $W_{c,\lambda}$ at roughly the same place, whereas the second crossing is a lot smoother. Another thing to note is that the curve created by the first crossing as a function of $\lambda$ seems to be cubic in $\lambda$, whereas the curve for second crossing seems to be quadratic.

\begin{figure}[t]
    \centering
    \includegraphics[width=0.5\textwidth]{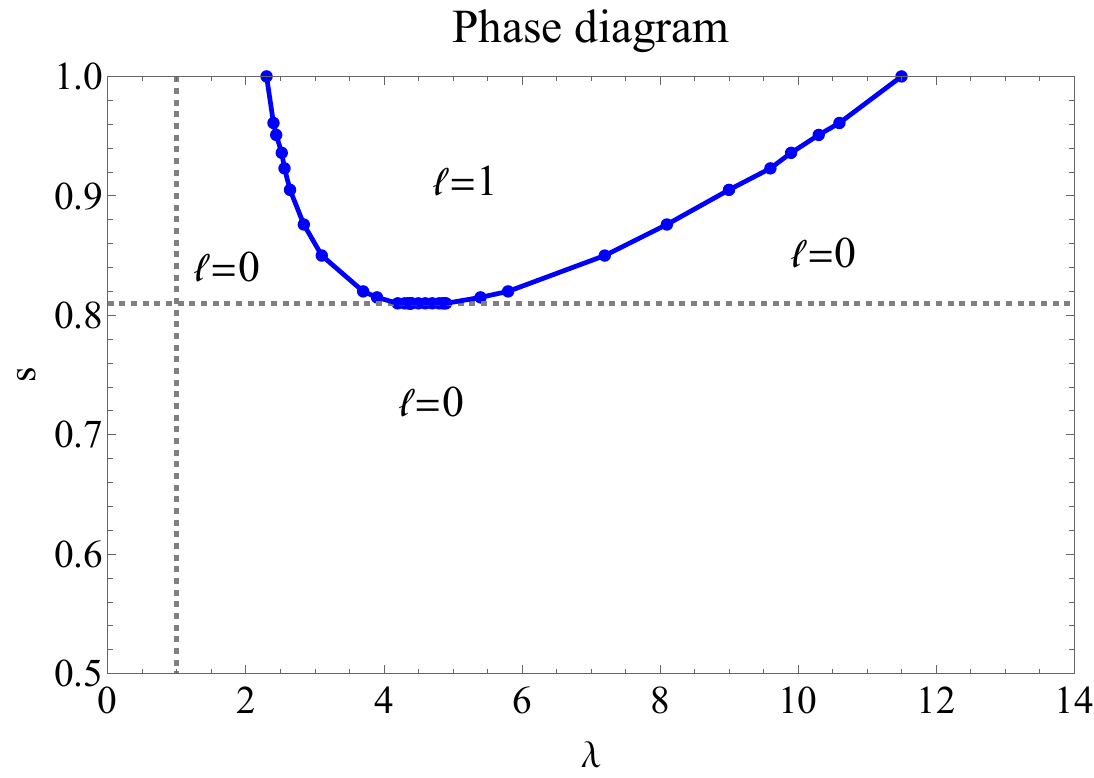}
    \caption{The phase diagram of  $E_{s,\lambda}^{\rm HF}$, showing in each region the angular momentum $\ell$ corresponding to the lowest energy.}
    \label{fig:phase}
\end{figure}

In Fig.~\ref{fig:phase} we report the resulting phase diagram in the $\lambda,s$ plane, showing the regions in which each $\ell$ gives the lowest energy. We observe two distinct regions: the first, with $0.5\leq s\leq 0.810$ ($0.1063\leq w\leq0.8937$), has no crossing of states, with $\ell=0$ being the ground state at all $\lambda\ge 0$, which means that there are also no discontinuities in $W_{c,\lambda}$.  The second region, $s>0.810$, has two crossings of states, with an intermediate $\lambda$ range in which the $\ell=1$ channel gives the lowest energy. The $\ell >1$ channels have been found to always have higher energy in the physical range $s\in [\frac12,1]$.

Similarly to constraining the spin, one can also follow the AC along the constrained $\ell=0$ channel. This removes the crossings of states that introduce discontinuities in $W_{c,\lambda}$, see, as an example, the right panels of Fig.~\ref{fig:EWgraphs}. Since the crossings always happen for $\lambda>1$, interpolating between the large and small $\lambda$ limits along the $\ell=0$ curve allows us to make approximations of $W_{c,\lambda}$ without affecting the resulting correlation energy.  The only exception remains $s=1$, which still has a discontinuity, this time at $\lambda=2.5$ (see the left panels of Fig. \ref{fig:EWgraphs}), due to a crossing between the flat 1s curve $E_{s=1,\lambda}^{\rm HF}=-0.5$ and the second $\ell=0$ state, with a radial node. However, in all the other cases the transitions are smooth, meaning that continuous curves for $0.5\leq s<1$ can be obtained when constraining $\ell=0$.

\section{The $\lambda\to\infty$ coefficients for the H atom}
\label{sec:HatomLargeLambda}
In this section we compute the spin-dependence of the first three leading terms of the large-$\lambda$ expansion of the H atom MPAC studied in the previous Sec.~\ref{sec:MPACHatom}. We will then show in the next Sec.~\ref{sec:manyelectrons} that, similarly to the closed-shell case,\cite{DaaGroVucMusKooSeiGieGor-JCP-20} these coefficients can be used in the general many-electron case.

We start from the Euler-Langrange equation \eqref{EulerLagrange-lambda}. As $\lambda\to\infty$, the effect of the term $-\lambda\,v^{\rm H}_s(r)\,u(r)$ becomes dominant \cite{DaaGroVucMusKooSeiGieGor-JCP-20}: despite the presence of the quantum-mechanical kinetic-energy term and of the exchange operator, the solution $u(r)$ will for $\lambda\to\infty$ concentrate indefinitely at the minimum of $-v^{\rm H}_s(\rv)$ at $\rv={\bf 0}$, implying
\begin{equation}
E_{s,\lambda}(\ell)\;=\;-\lambda\,v^{\rm H}_s(0)\,+\,O\big(\lambda^{\frac12}\big).%\,+\,\delta E_{w,\lambda}(\ell)
\label{concentration}\end{equation}
To see this explicitly, Eq.~\eqref{E-PTlambdaInfty} below, we expand \cite{DaaGroVucMusKooSeiGieGor-JCP-20} in Eq.~\eqref{EulerLagrange-lambda}
\begin{align}
v_s^{\rm H}(r) & = v_s^{\rm H}(0)-\frac{R_s(0)^2}6\,r^2+Z\,\frac{R_s(0)^2}6\,r^3+O(r^4),\\
R_s(r) & = R_s(0)\big(1\,-\,Z\,r\big)+O(r^2),
\end{align}
where we have used the cusp condition $R'_s(0)=-Z R_s(0)$ (see subsection II.B above).
%{\color{red} Notice, that $v_w^{\rm H}(r)$ would have no term $O(r^3)$, if $R_w(r)$ did not have a cusp, $R'_w(0)=0$.}
We emphasize that $v_s^{\rm H}(r)$, in the hypothetic case $R'_s(0)=0$, would have no term $O(r^3)$.
In terms of the scaled coordinate
\begin{equation}
t=\lambda^{\frac14}r,
\nonumber\end{equation}
writing $u(r)=u(\lambda^{-\frac14}t)=\lambda^{\frac18}\bar{u}(t)$, with $\int_0^{\infty}{\rm d}t\,\bar{u}(t)^2=1$, Eq.~\eqref{EulerLagrange-lambda} takes the form
\begin{equation}
E_{s,\lambda}(\ell)\,\bar{u}(t)=\Big[\!\!-v_s^{\rm H}(0)\,\lambda+\hat{H}_{\frac12}\,\lambda^{\frac12}+Z\hat{h}_{\frac14}\,\lambda^{\frac14}+O(\lambda^0)\Big]\bar{u}(t)
\nonumber\end{equation}
with $\lambda$-independent operators $\hat{H}_{\frac12}$ and $\hat{H}_{\frac14}=Z\hat{h}_{\frac14}$ given explicitly in Ref.~\onlinecite{DaaGroVucMusKooSeiGieGor-JCP-20}. Here $\hat{H}_{\frac12}$ and $\hat{h}_{\frac14}$ neither depend on $Z$ nor on $v_s^{\rm H}(0)$, but still on $s$, $\ell$, and (via the above expansions) also on $R_s(0)$. Writing
\begin{equation}
E_{s,\lambda}(\ell)=-v_s^{\rm H}(0)\,\lambda+\epsilon_{\frac12}\,\lambda^{\frac12}+\epsilon_{\frac14}\,\lambda^{\frac14}+O(\lambda^{0}),
\label{E-PTlambdaInfty}\end{equation}
we see that $\epsilon_{\frac12}$ is the zero-order eigenvalue $\epsilon^{(0)}$ in the perturbation expansion for the operator $\hat{H}_{\frac12}+Z\hat{h}_{\frac14}\lambda^{-\frac14}$,
\begin{equation}
\hat{H}_{\frac12}\bar{u}^{(0)}(t)=\epsilon_{\frac12}\bar{u}^{(0)}(t),
\label{eigenEq-0th}\end{equation}
while $\epsilon_{\frac14}$ is the corresponding first-order correction $\epsilon^{(1)}$,
\begin{equation}
\epsilon_{\frac14}=\big\langle\bar{u}^{(0)}\big|Z\hat{h}_{\frac14}\big|\bar{u}^{(0)}\big\rangle.
\label{eigenEq-1st}\end{equation}
Via the operators $\hat{H}_{\frac12}$ and $\hat{h}_{\frac14}$, the values $\epsilon_{\frac12}$ and $\epsilon_{\frac14}$ are depending on $R_s(0)=\sqrt{4\pi\rho_s^{\rm HF}(0)}$. This dependence is easily revealed,
\begin{align}
\epsilon_{\frac12} & = \tilde{\epsilon}_{\frac12}\,R_s(0) \nonumber\\
\epsilon_{\frac14} & = \tilde{\epsilon}_{\frac14}\,Z\sqrt{R_s(0)}
\nonumber\end{align}
when we write $\bar{u}^{(0)}(t)=\bar{u}^{(0)}(\frac{p}C)=\sqrt{C}u_{\frac12}(p)$, with $C=\sqrt{R_s(0)}$ and $\int_0^{\infty}{\rm d}p\,u_{\frac12}(p)^2=1$.
Then, Eq.~\eqref{eigenEq-0th} becomes a universal equation \cite{DaaGroVucMusKooSeiGieGor-JCP-20} with two parameters $\ell$ and $s$,
\begin{align}
& -\frac12\,u''_{\frac12}(p)
+\left(\frac{\ell(\ell+1)}{2p^2}+\frac{p^2}6\right)u_{\frac12}(p)\nonumber\\
& + \frac{s}{2\ell+1}\bigg[\frac{1}{p^{\ell}}\int_0^p{\rm d}p'\,(p')^{\ell+1}\,u_{\frac12}(p')\nonumber\\
& \qquad + p^{\ell+1}\int_p^\infty\frac{{\rm d}p'}{(p')^{\ell}}\,u_{\frac12}(p')\bigg] = \tilde{\epsilon}_{\frac12}\,u_{\frac12}(p).\label{eq:HFlbl0eps12}
\end{align}
 Numerically, the eigenvalue $\tilde{\epsilon}_{\frac12}=\tilde{\epsilon}_{\frac12}(\ell,s)$ is always lowest for $\ell=0$, independently of $s$, confirming the results of the phase diagram of Fig.~\ref{fig:phase}. Therefore, for the determination of the large-$\lambda$ coefficients we can set $\ell=0$ everywhere. Then Eq.~\eqref{eigenEq-1st} reads
%As this leading coefficient $\tilde{\epsilon}_{\frac12}$ (numerically) is always lowest for $\ell=0$, we can for $\lambda\gg1$ generally choose $\ell=0$. Then Eq.~\eqref{eigenEq-1st} reads
\begin{align}
\tilde{\epsilon}_{\frac14}& = \int_0^{\infty}{\rm d}p\,u_{\frac12}(p)\,\hat{h}_{\frac14}u_{\frac12}(p)\label{eq:HFlbl0eps14}\\
& = -\int_0^\infty dp\,u_{\frac{1}{2}}(p)\,\bigg\{\left(\frac{1}{p}+\frac{p^3}{6}\right) u_{\frac{1}{2}}(p)\nonumber\\
& +\,2s\,\bigg[p\int_0^p {\rm d}p'\, p'\, u_{\frac{1}{2}}(p')+\int_0^p {\rm d}p'\, (p')^2\, u_{\frac{1}{2}}(p')\bigg]\bigg\}.
\nonumber\end{align}
We then see that $\tilde{\epsilon}_{\frac12}$ and $\tilde{\epsilon}_{\frac14}$ for $\ell=0$ are pure functions of $s$, see Figs.~\ref{fig:eps12}
and \ref{fig:eps14}.

In summary, Eq.~\eqref{E-PTlambdaInfty} yields for $W_{{\rm c},s,\lambda} = \frac{{\rm d}E_{s,\lambda}}{{\rm d}\lambda}+(1-s)\,U[\rho_s^{\rm HF}]$ for large $\lambda\to\infty$ (when the minimizer in Eq.~\eqref{minCond3B} is always $\ell=0$), the expansion of Eq.~\eqref{eq:finalexp}
\begin{equation}
W_{{\rm c},s,\lambda\to\infty} = W_{{\rm c},\infty}(s) +\frac{W_{\frac12}(s)}{\sqrt{\lambda}}+\frac{W_{\frac34}(s)}{\lambda^{\frac34}}+...
\nonumber\end{equation}
where the coefficients now depend on the weight parameter $s=1-2w(1-w)$,	 
\begin{align}
W_{{\rm c},\infty}(s) & = -v_s^{\rm H}(0)+(1-s)\,U[\rho_s^{\rm HF}] \nonumber\\
W_{\frac12}(s) & = \frac{\tilde{\epsilon}_{\frac12}(s)}{2}\sqrt{4\pi\rho_s^{\rm HF}(0)}\nonumber\\
W_{\frac34}(s) & = Z\frac{\tilde{\epsilon}_{\frac14}(s)}{4}\sqrt[4]{4\pi\rho_s^{\rm HF}(0)}.
\nonumber\end{align}

\subsection{Results: the functions $\tilde{\epsilon}_{\frac12}(s)$ and $\tilde{\epsilon}_{\frac14}(s)$}
%which contains all terms that contribute at order $\lambda=1/2$, which are the (angular) kinetic energy, Hartree potential and the exchange potential.
To compute $\tilde{\epsilon}_{\frac12}(s)$ and $\tilde{\epsilon}_{\frac14}(s)$ we set $\ell=0$ in 
Eq.~\eqref{eq:HFlbl0eps12} and expand $u_{\frac{1}{2}}(p)$ on the basis of the quantum isotropic harmomic oscilator (IHO) problem that arises if we set $s=0$ in Eq.~\eqref{eq:HFlbl0eps12}, which has frequency $\omega=\frac{1}{\sqrt{3}}$,
and  energies given by  $\frac{\omega(3+4n)}{2}$,
\begin{align}\label{eq:HFLargeLB}
 u_{\frac{1}{2}}(p) & = p\,\sum_n c_n(s) \xi_{n}(p) \\
 \xi_{n}(p) & = \mathcal{N}\,\exp \left(-\frac{p^2}{2
   \sqrt{3}}\right) L_n^{\frac{1}{2}}\left(\frac{p^2}{\sqrt{3}}\right) \nonumber \\
   \mathcal{N} & = \sqrt{\frac{\sqrt{\frac{1}{4 \pi  \sqrt{3}^3}} 2^{n+3} n!}{(2 n+1)\text{!!}}}, \nonumber 
%   \left(-\frac{1}{2}\frac{d^2}{dp^2}+\frac{l(l+1)}{2p^2}+\frac{1}{6}p^2\right) p\psi_{n}(p)& =\left(2n+l+\frac{3}{2}\right)p\psi_{n}(p)
\end{align}
with $L_{n}^{\frac{1}{2}}$ being generalized Laguerre polynomials. 

\begin{figure}[t]
    \centering
\includegraphics[width=0.5\textwidth]{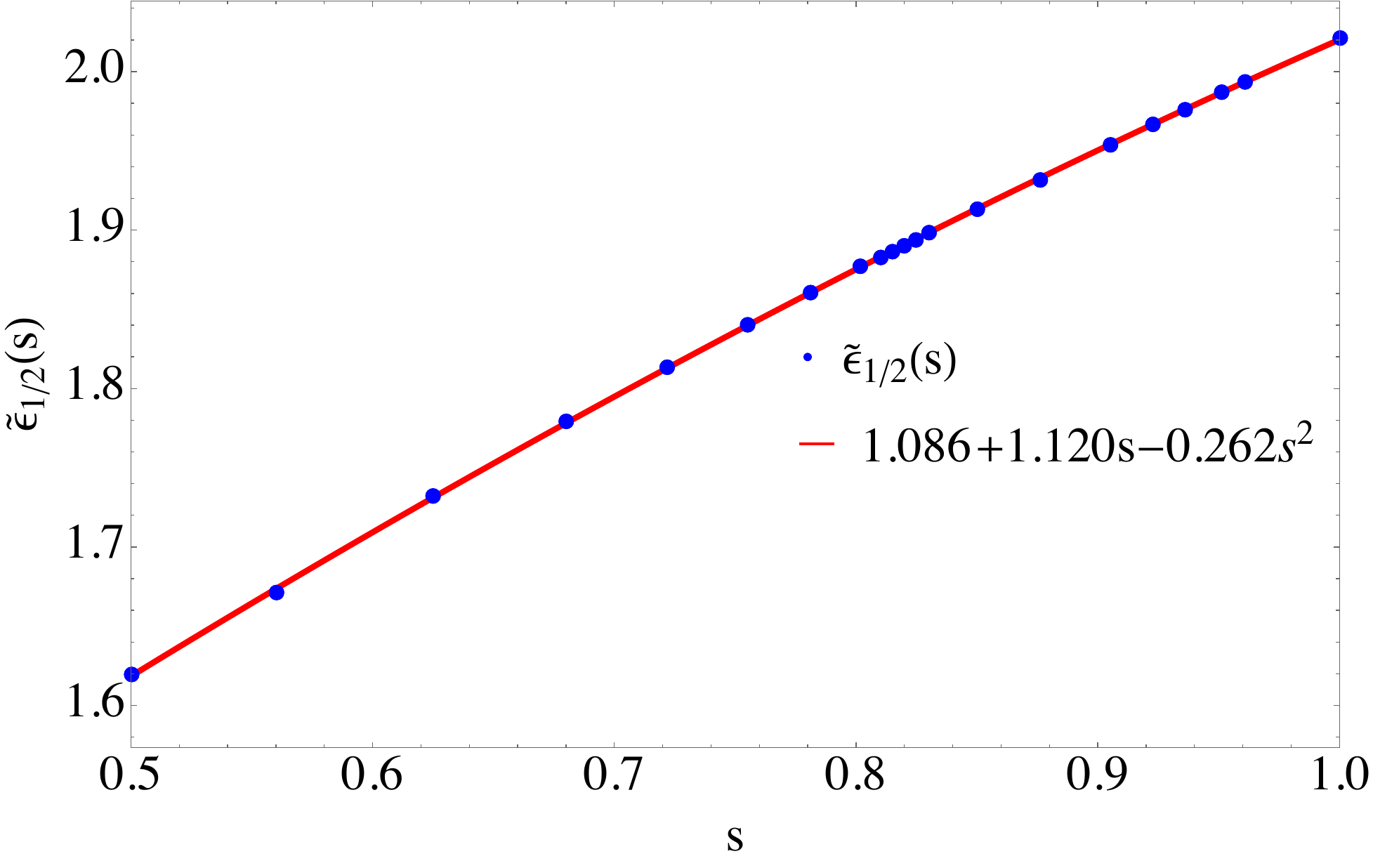}
 \caption{The dependence of $\tilde{\epsilon}_\frac{1}{2}$ on $s$ between $0.5$ and $1$ and a quadratic fit.}
    \label{fig:eps12}
\end{figure}

\begin{figure}[t]
    \centering  \includegraphics[width=0.5\textwidth]{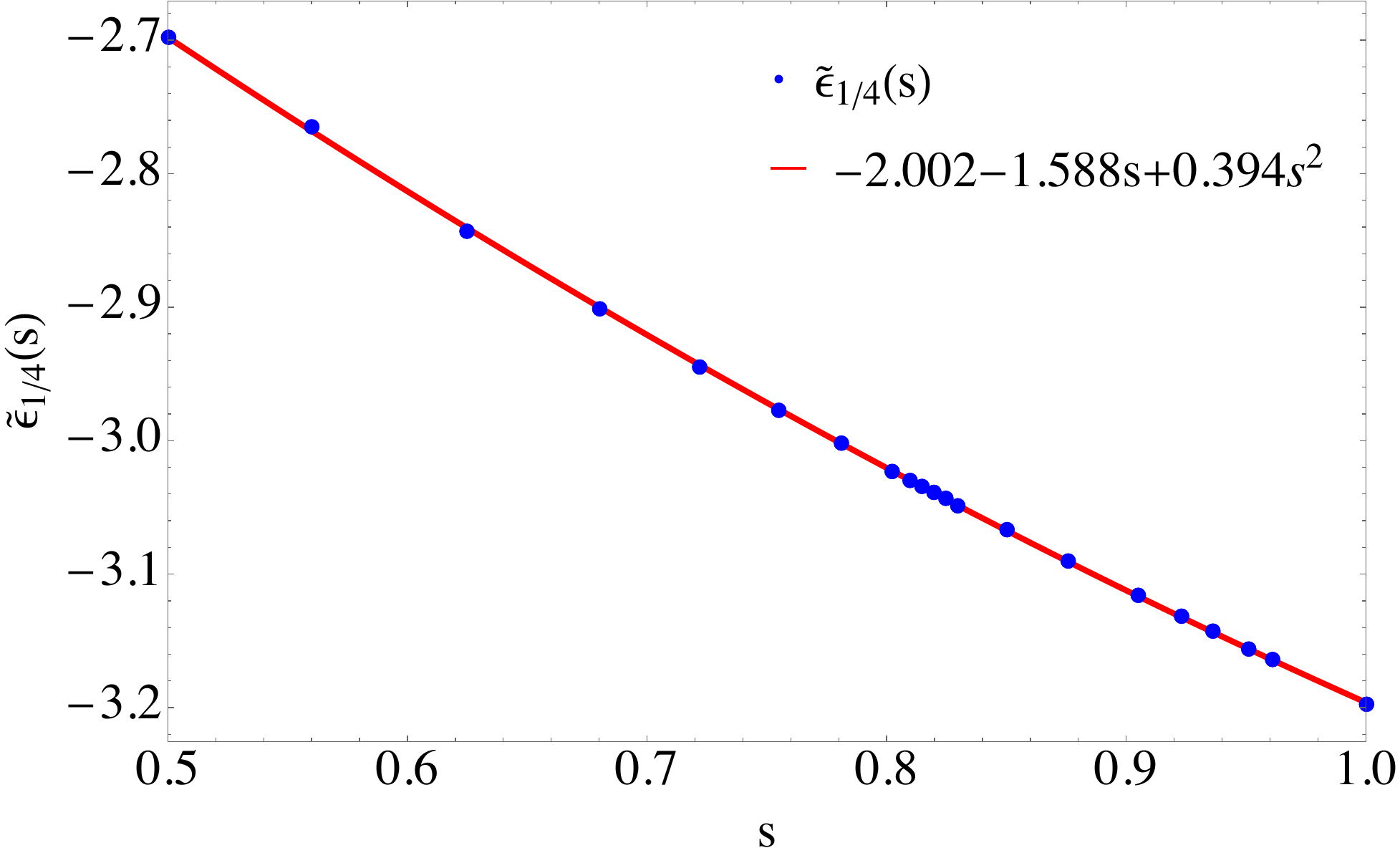}
    \caption{The dependence of $\tilde{\epsilon}_\frac{1}{4}$ on $s$ between $0.5$ and $1$ and a quadratic fit.}
    \label{fig:eps14}
\end{figure}

Numerical solutions of Eq.~\eqref{eq:HFlbl0eps12} with $\ell=0$ for different values of $s\in [\frac12,1]$ have thus been obtained both by using the IHO basis set expansion of equation \eqref{eq:HFLargeLB}, which converges very fast, and by using the spectral renormalization method, finding perfect agreement for $\tilde{\epsilon}_{\frac12}(s)$ and $\tilde{\epsilon}_{\frac14}(s)$, which are shown, respectively, in Figs.~\ref{fig:eps12} and \ref{fig:eps14}.

%In Figure \ref{fig:eps12}, an interestingly (almost) linear behaviour of $\epsilon_{\frac{1}{2}}$ for $s$ is shown. To understand this, one needs to notice that the only direct $s$ dependence comes for the exchange integrals, meaning that we can rewrite the equation in following way,
%\begin{equation}
%    \epsilon_{\frac{1}{2}}=\left<u_{\frac{1}{2}}(p)\left|\hat{h}(p)\right|u_{\frac{1}{2}}(p)\right>+s\left<u_{\frac{1}{2}}(p)\left|\hat{k}(p)\right|u_{\frac{1}{2}}(p)\right>,
%\end{equation}
%with
%\begin{equation}
%    \hat{h}(p)u_{\frac{1}{2}}(p)=-\frac{1}{2}u_{\frac{1}{2}}''(p)+\frac{1}{6}p^2 u_{\frac{1}{2}}(p)
%\end{equation}
%and 
%\begin{equation}
%    \hat{k}(p)u_{\frac{1}{2}}(q)=\int_{0}^{p}dq~q u_{\frac{1}{2}}(q)+p\int_{p}^{\infty}dq~ u_{\frac{1}{2}}(q).
%\end{equation}
%This means that $\epsilon_{\frac{1}{2}}$ will have mostly a linear behaviour with $s$, but with small higher contributions coming from the indirect dependence on the minimizing wave function $u_{\frac{1}{2}}$. The same is true for $\epsilon_{\frac{1}{4}}$, which also has a direct linear dependence on $s$ in equation \eqref{eq:HFlbl0eps14}, and is shown in Fig \ref{fig:eps14}. 
In both figures $\tilde{\epsilon}_{\frac{1}{2}}(s)$ and $\tilde{\epsilon}_{\frac{1}{4}}(s)$ are accompanied by a quadratic fit, which proves to accurately interpolate the results and can be used for all practical purposes with Max Absolute Deviations of 0.0026 and 0.0032 respectively.

In appendix~\ref{app:proof} we report an intriguing curiosity regarding Eq.~\eqref{eq:HFlbl0eps12}: it has closed-form solutions for special pairs of $s$ and $\ell$ values. It was already noticed in Ref.~\onlinecite{DaaGroVucMusKooSeiGieGor-JCP-20} that the case $s=1,\ell=0$ has a simple closed-form solution, and in appendix~\ref{app:proof} we investigate the structure of Eq.~\eqref{eq:HFlbl0eps12} further, finding an infinite set of such solutions. Unfortunately, they all appear at $s>1$ and thus have no physical meaning in the context we are analyzing.

\section{General many-electron case}\label{sec:manyelectrons}
We show in this section that the $\lambda\to\infty$ results for the H atom at $s\in [\frac{1}{2},1]$ provide a variational estimate for the strong-coupling leading terms of the general many-electron case, in terms of functionals of the HF $\alpha$-spin and $\beta$-spin densities.

The derivation is a generalization of the one for the spin-unpolarized (closed-shell) case:\cite{DaaGroVucMusKooSeiGieGor-JCP-20} we start from a variational ansatz for the wavefunction $\Psi_{\lambda\to\infty}$ that minimizes the Hamiltonian \eqref{HamHFlambdaN} when $\lambda\to\infty$.  We use a simple Hartree product of localized orbitals around each minimizing position $\rv^{\rm min}_i$ of Eq.~\eqref{eq:EelDef}, with each spin in a superposition (again, for all the expectation values we need to consider in the derivation, it is the same if we use an ensemble for the spin part instead of a superposition). Anti-symmetry of the wavefunction can be neglected as it contributes to orders $\sim e^{-\sqrt{\lambda}}$ to the energy.\cite{GroKooGieSeiCohMorGor-JCTC-17,SeiGiaVucFabGor-JCP-18,DaaGroVucMusKooSeiGieGor-JCP-20} The Hartree product then reads
\begin{align}\label{eq:PsiT2}
	& \Psi_\lambda^{h}(\xv_1,\dots,\xv_N)  = \\
 & =\prod_{i=1}^N\loc_{i,\lambda}(|\rv_i-\rv^{\rm min}_i|)\left(\sqrt{1-q_i}\,|\alpha(i)\rangle+\sqrt{q_i}\,|\beta(i)\rangle\right),
\end{align}
where 
\begin{equation}\label{eq:loc1}
	\loc_{i,\lambda}(r)=\lambda^{\frac{3n}{2}}\loc_i(\lambda^n\,r),
\end{equation}
$\loc_i(t)$ is a localised, normalised, 3D spherical function,
\begin{equation}
	\int d{\bf t}\, \loc_i^2(t)=1,
\end{equation}
which needs to be determined variationally, $0\le q_i\le 1$, and we will set later $n=\frac{1}{4}$.\cite{SeiGiaVucFabGor-JCP-18,DaaGroVucMusKooSeiGieGor-JCP-20} In other words, we know that the wavefunction squared $|\Psi_{\lambda\to\infty}|^2$ tends to a product of delta functions centered around the minimizing positions  $\rv^{\rm min}_i$, and with our ansatz we seek the best variational spherical representation of the delta function that minimizes the next leading term. This ansatz does not take into account the coupling between the localized states and their anisotropy. As such, it can only provide a variational upper bound for the $\lambda\to\infty$ functionals. The same kind of approximation was used by Wigner\cite{Wig-TFS-38} to compute the zero-point energy in the low-density electron gas, yielding an error of $\sim 12\%$ with respect to the full coupled exact solution,\cite{AlvBenEvaBer-PRB-21} which could provide an indication on the tightness of the upper bound we provide.

We thus evaluate the expectation of the Hamiltonian $\hat{H}_{\lambda}^{\rm HF}$ of Eq.~\eqref{HamHFlambdaN} on $\Psi_\lambda^{h}$, and we retain only the leading orders at $\lambda\to\infty$. The kinetic energy and $\hat{V}_{ee}-\hat{J}$ are spin-independent: their expectation values at large $\lambda$ is the same as for the closed-shell case considered in Ref.~\onlinecite{DaaGroVucMusKooSeiGieGor-JCP-20}, which we report here again for completeness,
\begin{align}\label{eq:TonL}
	\langle \Psi_\lambda^{h}|\hat{T}|\Psi_\lambda^{h}\rangle=\frac{\lambda^{2n}}{2}\sum_{i=1}^N\int d{\bf t}|\nabla\loc_i(t)|^2,
	%\equiv \lambda^{2n}\sum_{i=1}^N \tilde{T}_{\frac{1}{2}}[\loc_i]
\end{align}
and 
\begin{align}
& \langle \Psi_\lambda^{h}|\hat{V}_{ee}-\hat{J}|\Psi_\lambda^{h}\rangle-C=\nonumber \\
& \lambda^{-2n}\sum_{i=1}^N 4\pi\,\rho^{\rm HF}(\rv_i^{\rm min})
\int d\tv\, \frac{t^2}{6}\,\loc_i^2(t) +O(\lambda^{-3n})
, \label{eq:Vee-JonL}
\end{align}
where $C=E_{\rm el}[\rho^{\rm HF}]-U[\rho^{\rm HF}]$.

The expectation of $\hat{K}$ is the only part that changes with respect to the closed-shell case. The kernel of $\hat{K}$ for a general open-shell system with $N_\alpha$ spin-up electrons and $N_\beta=N-N_\alpha$ spin-down electrons reads
\begin{widetext}
\begin{align} k(\xv,\xv')=\frac{\sum_{a=1}^{N_\alpha}\phi_{a,\alpha}^*(\rv')\phi_{a,\alpha}(\rv)\langle\sigma|\alpha\rangle\langle\alpha|\sigma'\rangle+\sum_{a=1}^{N_\beta}\phi_{a,\beta}^*(\rv')\phi_{a,\beta}(\rv)\langle\sigma|\beta\rangle\langle\beta|\sigma'\rangle}{|\rv-\rv'|},
\end{align}
and its expectation on $\Psi_\lambda^{h}$ at large $\lambda$ is then
\begin{align}
	& \langle \Psi_\lambda^{h}|\hat{K}|\Psi_\lambda^{h}\rangle  =\sum_{i=1}^N\int d\rv\int d\rv'\frac{\loc_{i,\lambda}(|\rv-\rv^{\rm min}_i|)\loc_{i,\lambda}(|\rv'-\rv^{\rm min}_i|)}{|\rv-\rv'|}\left((1-q_i)\sum_{a=1}^{N_\alpha}\phi_{a,\alpha}^*(\rv')\phi_{a,\alpha}(\rv)+q_i\sum_{a=1}^{N_\beta}\phi_{a,\beta}^*(\rv')\phi_{a,\beta}(\rv)\right) \nonumber \\
	& =\lambda^{-2n}\sum_{i=1}^N\int d\tv\int d\tv'\frac{\loc_i(t)\loc_i(t')}{|\tv-\tv'|}\biggl((1-q_i)\,\underbrace{\sum_{a=1}^{N_\alpha}|\phi_{a,\alpha}(\rv_i^{\rm min})|^2}_{=\rho_\alpha^{\rm HF}(\rv_i^{\rm min})}+q_i \,\underbrace{\sum_{a=1}^{N_\beta}|\phi_{a,\beta}(\rv_i^{\rm min})|^2}_{=\rho_\beta^{\rm HF}(\rv_i^{\rm min})}\biggr)+O(\lambda^{-3n}) \nonumber \\
&  = \lambda^{-2n}  \sum_{i=1}^N \rho^{\rm HF}(\rv_i^{\rm min})\biggl((1-q_i)\,\frac{(1+\zeta^{\rm HF}(\rv_i^{\rm min}))}{2}+q_i \,\frac{(1-\zeta^{\rm HF}(\rv_i^{\rm min}))}{2}\biggr)\int d\tv\int d\tv'\frac{\loc_i(t)\loc_i(t')}{|\tv-\tv'|}+O(\lambda^{-3n}), \label{eq:KonL}
\end{align}
\end{widetext}
where we have expanded the HF spin-orbitals $\phi_{a,\sigma}$ in scaled coordinates
$\tv_i=\lambda^{n}(\rv_i-\rv_i^{\rm min})$ at large $\lambda$,
\begin{equation}
	\phi_{a,\sigma}(\lambda^{-n}\tv_i+\rv_i^{\rm min})=\phi_{a,\sigma}(\rv_i^{\rm min})+O(\lambda^{-n}),
\end{equation}
and we have introduced the usual spin-polarization parameter for the HF $\alpha$-spin and $\beta$-spin densities,
\begin{align}
\label{eq:zeta}
    \zeta^{\rm HF}(\rv)=\frac{\rho_\alpha^{\rm HF}(\rv)-\rho_\beta^{\rm HF}(\rv)}{\rho^{\rm HF}(\rv)}.
\end{align}
Now we see that the local spin polarization at the minimizing positions plays exactly the same role as $w$ in our derivation for the H atom, with $w=\frac{1-\zeta}{2}$ and $1-w=\frac{1+\zeta}{2}$. If we want to use the $\lambda\to\infty$ expansion to build interpolations we need to forbid spin flip (to keep $W_{c,\lambda}$ continuous) and thus set
\begin{align}
    q_i=\frac{1-\zeta^{\rm HF}(\rv_i^{\rm min})}{2} \Rightarrow 1-q_i=\frac{1+\zeta^{\rm HF}(\rv_i^{\rm min})}{2}.
\end{align}
With this choice, the leading ($\lambda^{-2n})$ term of Eq.~\eqref{eq:KonL} becomes
\begin{align}
\label{eq:leadK}
    \sum_{i=1}^N \rho^{\rm HF}(\rv_i^{\rm min})\frac{1+\zeta^{\rm HF}(\rv_i^{\rm min})^2}{2}\int d\tv\int d\tv'\frac{\loc_i(t)\loc_i(t')}{|\tv-\tv'|}.
\end{align}
We thus see that if we set
\begin{align}
\label{eq:szeta}
    s(\rv)=\frac{1+\zeta^{\rm HF}(\rv)^2}{2},
\end{align}
(which varies between $\frac{1}{2}$ and 1, exactly as in Sec.~\ref{sec:MPACHatom}), we insert Eqs.~\eqref{eq:TonL}, \eqref{eq:Vee-JonL} and  \eqref{eq:leadK} in the expectation of $\hat{H}_{\lambda}^{\rm HF}$ of Eq.~\eqref{HamHFlambdaN} and set $n=1/4$, we obtain, neglecting orders $\lambda^{1/4}$ and lower,
\begin{align}
\langle \Psi_\lambda^{h}|\hat{H}_{\lambda}^{\rm HF}|\Psi_\lambda^{h}\rangle-\lambda C=\lambda^{1/2}\sum_{i=1}^N \tilde{E}_{s(\rv_i^{\rm min})}(\rho^{\rm HF}(\rv_i^{\rm min}))[\loc_i] ,
\end{align}
where
\begin{align}
	\tilde{E}_{s}(\rho)[\loc] & =\frac{1}{2}\int d{\bf t}|\nabla\loc(t)|^2+4\pi\,\rho\,
\int d\tv\, \frac{t^2}{6}\,\loc^2(t) \nonumber \\
& +\rho\,s\,
	\int d\tv\int d\tv'\frac{\loc(t)\loc(t')}{|\tv-\tv'|}.
\end{align}
Exactly as in the closed-shell case,\cite{DaaGroVucMusKooSeiGieGor-JCP-20} varying $\tilde{E}_{s}(\rho)[\loc]$ with respect to $\loc$ (keeping the normalisation constraint), switching to the function $ u_{\frac{1}{2}}(t)=\sqrt{4\pi}\,t\,\loc(t)$, and introducing the scaled variable $p=(4\pi\,\rho)^{1/4} t$, we obtain Eq.~\eqref{eq:HFlbl0eps12} with $\ell=0$. This means that the best possible spherical variational ansatz for $\loc_i$ is the same as the one we found for the $s$-dependent H atom in Sec.~\ref{sec:MPACHatom}, around each equilibrium position $\rv_i^{\rm min}$, yielding the variational estimate
\begin{align}
W_{\frac{1}{2}}[\rho_\alpha^{\rm HF},\rho_\beta^{\rm HF}] \approx \frac{1}{2}\sum_{i=1}^N\tilde{\epsilon}_{\frac{1}{2}}(s(\rv_i^{\rm min}))\sqrt{4\pi\rho^{\rm HF}(\rv_i^{\rm min})},\label{eq:W12fromH}
\end{align}
where we can use for $\tilde{\epsilon}_{\frac{1}{2}}(s)$ the quadratic fit of Fig.~\ref{fig:eps12}, and $s(\rv)$ is a functional of the HF $\alpha$-spin and $\beta$-spin densities via Eqs.~\eqref{eq:zeta} and \eqref{eq:szeta}.

The next leading order works exactly as in the closed-shell case.\cite{DaaGroVucMusKooSeiGieGor-JCP-20} Its generalization to the open-shell case is then 
\begin{align}
W_{\frac{3}{4}}[\rho_\alpha^{\rm HF},\rho_\beta^{\rm HF}] \approx\frac{1}{4} \sum_{i=1}^{I_{\rm nuc}}Z_{k_i}\tilde{\epsilon}_{\frac{1}{4}}(s(\Rv_{k_i}))\,\sqrt[4]{4\pi\rho^{\rm HF}(\Rv_{k_i})}
\end{align}
where the sum runs only over the nuclear positions that coincide with minimizing electronic positions $\rv^{\rm min}_i$, as in Eq.~\eqref{eq:min14}. For the function $\tilde{\epsilon}_{\frac{1}{4}}(s)$ we can use the quadratic fit of Fig.~\ref{fig:eps14}.

We notice that these variational estimates are strictly valid for the restricted open shell HF case only. For the unrestricted case, there would be an additional dependence on the $\alpha$-spin and $\beta$-spin densities entering when we solve the $s$-dependent equation for the H atom, as the pair $\phi_{s,\alpha}({\bf 0})$ and  $\phi_{\beta,\alpha}({\bf 0})$ appear in those equations, and the resulting functions $\tilde{\epsilon}_{\frac{1}{2}}(s)$ and $\tilde{\epsilon}_{\frac{1}{4}}(s)$ will be slightly different. However, we may expect these effects to be much smaller than the main $s$-dependence studied here (see, e.g., Fig.~1 of Ref.~\onlinecite{BurMarDaaGorLoo-JCP-21}).

\section{Conclusions and perspectives}
\label{sec:conc}
We have extended the results\cite{DaaGroVucMusKooSeiGieGor-JCP-20} for the large-coupling limit of the adiabatic connection that has as small-coupling expansion the M{\o}ller-Plesset series to the open-shell case. We first studied the paradigmatic case of the H atom, revealing an interesting phase diagram (Fig.~\ref{fig:phase}), and we then showed that the results for the H atom at large coupling strength can be used for the general many-electron open shell case (Sec.~\ref{sec:manyelectrons}), yielding functionals of the HF $\alpha$-spin and $\beta$-spin densities.

Using these results, we plan to extend the construction of MPAC functionals\cite{DaaFabDelGorVuc-JPCL-21,DaaKooPetFabDelGorVuc-JPCL-23} to open shell systems, either by developing generalized gradient approximations for the leading term functionals, as done for the closed-shell case,\cite{DaaKooGroSeiGor-JCTC-22} or by using inequalities and relationships with the DFT AC case.\cite{DaaFabDelGorVuc-JPCL-21,DaaKooPetFabDelGorVuc-JPCL-23}

\section*{Acknowledgements}
It is a pleasure to dedicate this paper to Gustavo Scuseria, whose deep work and understanding of strong correlation in many-electron systems has been a continuous source of inspiration. We also want to thank Klaas Giesbertz for his suggestions.
Financial support from the Netherlands Organisation for Scientific Research (NWO) under Vici grant 724.017.001 is acknowledged.

\section*{TOC graphic}
\begin{figure}[h]
    \centering
    \includegraphics[width=0.25\textwidth]{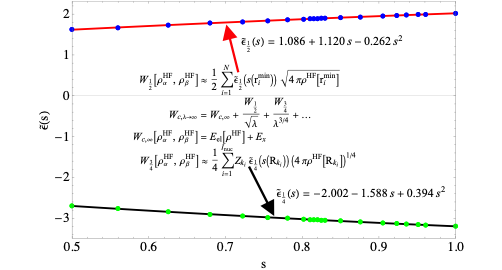}
    \label{fig:toc}
\end{figure}

%\section*{Data Availability Statement}

\appendix
\section{Closed-form solutions for special values of $s$ at $\ell=0$}\label{app:proof}
We start by noticing that the energy, $\tilde{\epsilon}_{\frac{1}{2}}$, for $s=1$ and $\ell=0$ is exactly equal to the energy of the first excited state of the isotropic harmonic oscillator, which arises by setting $s=0$ in Eq.~\eqref{eq:HFlbl0eps12} ($\omega=1/\sqrt{3}$). Furthermore, the first excited state of $s=1$ is again equal to the energy of the second excited state of $s=0$, and has as closed-form solution a linear combination of the ground state and the first two excited states. In summary, the energies of $s=1$ and $s=0$ are related and their structure is shown in the first three columns of table~\ref{tab:IHOstruc}. Notice that in order to yield an energy of $\frac{\omega(7+4n)}{2}$ for $s=1$, one needs to mix in all IHO's orbitals until $n$. 

One can find other values of $s$ that give closed-form solutions to Eq.~\eqref{eq:HFlbl0eps12}  at $\ell=0$. The next one, $s=\frac{10}{3}$, is degenerate with the second excited state of the IHO ($s=0$), $\tilde{\epsilon}_{\frac{1}{2}}=\frac{11}{2\sqrt{3}}$, and is again a linear combination of the ground state and the first two excited states at $s=0$. Closed-form ground- and excited-state energies for $s=7$, $s=12$ and $s=\frac{55}{3}$ can also be obtained, see Table~\ref{tab:IHOstruc}. One can find a formula for the values of these special $s_m$ and the corresponding energies $\tilde{\epsilon}_{\frac{1}{2}}(s_m)$,
\begin{equation}\label{eq:seq}
    s_m=\frac{m(2m+1)}{3}=\omega^2 m(2m+1),
\end{equation}
and
\begin{equation}\label{eq:epseq}
    \tilde{\epsilon}_{\frac{1}{2}}(s_m)=\frac{3+4n+4m}{2\sqrt{3}}=\frac{1}{2}\omega(3+4n+4m),
\end{equation}
with $m,n\in\mathbb{Z}$, and $n$ increasing the excitation rank. The corresponding wavefunctions for these values of $s$ are,
\begin{align}\nonumber
   u_{\frac{1}{2}}(p)&= \sum^{m-1}_{i=0}\sum^{n+i}_{j=0}c_{i,j}(m,n)\exp \left(-\frac{p^2}{2
   \sqrt{3}}\right) p L_{j}^{\frac{1}{2}}\left(\frac{p^2}{\sqrt{3}}\right)\\
   &=p\sum^{m-1}_{i=0}\sum^{n+i}_{j=0}c_{i,j}(m,n) \xi_{j}(p),
\end{align}
with 
\begin{align}
    c_{i,j}(m,n)&=\binom{n+i}{i}\binom{n+2m+1/2}{m-i}\binom{m+n+i-j-1}{n+i-j}\nonumber\\
    &\times(-1)^{n}\frac{2^{\frac{4m-3}{4}}}{3^{\frac{3+2m}{4}}\sqrt{\pi}}.\nonumber
\end{align}
%The proof for these relations is provided in the supplementary information. 

\begin{table}[t]
\caption{The structure of the closed-form solutions of equation \eqref{eq:HFlbl0eps12}, where the $c_{n}$'s of every combination of $\tilde{\epsilon}_{\frac{1}{2}}(s)$ and $s$ are different.}\label{tab:IHOstruc}
\begin{tabular}{l||l|l|l|l|l|l}
$\tilde{\epsilon}_{\frac{1}{2}}$ & $s=0$& $s=1$ & $s=\frac{10}{3}$ & $s=7$ & $s=12$ & $s=\frac{55}{3}$\\\hline\hline
$\frac{3}{2\sqrt{3}}$  & $c_{0}\xi_{0}$   &   &      &   &     & \\\hline
$\frac{7}{2\sqrt{3}}$   & $c_{1}\xi_{1}$  & $\sum_{n}^{1}c_{n}\xi_{n}$   &      &   &    & \\\hline
$\frac{11}{2\sqrt{3}}$ &  $c_{2}\xi_{2}$ &$\sum_{n}^{2}c_{n}\xi_{n}$   & $\sum_{n}^{2}c_{n}\xi_{n}$     &   &     & \\\hline
$\frac{15}{2\sqrt{3}}$  & $c_{3}\xi_{3}$  & $\sum_{n}^{3}c_{n}\xi_{n}$  & $\sum_{n}^{3}c_{n}\xi_{n}$      & $\sum_{n}^{3}c_{n}\xi_{n}$   &   &   \\\hline
$\frac{19}{2\sqrt{3}}$ & $c_{4}\xi_{4}$  & $\sum_{n}^{4}c_{n}\xi_{n}$  &  $\sum_{n}^{4}c_{n}\xi_{n}$     & $\sum_{n}^{4}c_{n}\xi_{n}$   &   $\sum_{n}^{4}c_{n}\xi_{n}$ &   \\\hline
$\frac{23}{2\sqrt{3}}$  &  $c_{5}\xi_{5}$ & $\sum_{n}^{5}c_{n}\xi_{n}$  & $\sum_{n}^{5}c_{n}\xi_{n}$      & $\sum_{n}^{5}c_{n}\xi_{n}$   &  $\sum_{n}^{5}c_{n}\xi_{n}$ &  $\sum_{n}^{5}c_{n}\xi_{n}$\\
\end{tabular}
\end{table}
Another way to look at these results is to start from the full 3D version of Eq.~\eqref{eq:HFlbl0eps12} before the partial wave expansion, namely
\begin{equation}\label{eq:3DFull}
-\frac{1}{2} \nabla^{2}\psi_\lambda(\mathbf{p})+\frac{1}{6}\mathbf{p}^2\psi_\lambda(\mathbf{p})+s \int \frac{\psi_\lambda(\mathbf{p}')}{|\mathbf{p}-\mathbf{p}'|}d\mathbf{p}'=E\psi_\lambda(\mathbf{p}).
\end{equation}
We can then expand  $\psi_\lambda(\mathbf{p})=\sum_{n}c_n\phi_n(\mathbf{p})$, with $\phi_n(\mathbf{p})$ the 3D IHO orbitals,
\begin{equation}\label{eq:3DExp}
    \sum_n \epsilon_n  c_n \phi_n(\mathbf{p}) + s \sum_n c_n \int\frac{\phi_n(\mathbf{p}')}{|\mathbf{p}-\mathbf{p}'|}d\mathbf{p}'=E\sum_n c_n\phi_n(\mathbf{r}).
\end{equation}
It can then be shown that the integral in the second term has the following expansion in terms of the $n-1$ states,
\begin{equation}\label{eq:erf}
    \int\frac{\phi_n(\mathbf{p}')}{|\mathbf{p}-\mathbf{p}'|}d\mathbf{p}'=\sum^{n-1}_{q} b_q^{(n)}\phi_{q}(\mathbf{p})+A_n \frac{\erf(\frac{p}{\sqrt{3}})}{p}
\end{equation}
Substituting equation \eqref{eq:erf} into equation \eqref{eq:3DExp}, immediately results in the following condition to equate the terms with the error functions,
\begin{equation}
    \sum_n^{m}A_n c_n=0,
\end{equation}
with $A_n = \int\phi_n (\mathbf{p}) d\mathbf{p}$ and $b_{q}^{(n)}$ being a constant that depends on $n$. The remaining terms result in a second condition that can be written as,
\begin{align}
    s \sum^{m}_{n=0} c_n \sum^{n-1}_{q=0}b_{q}^{(n)}\phi_{q}(\mathbf{p})=\sum^{m}_{n=0}c_n(E-\epsilon_n)\phi_{n}(\mathbf{p})
\end{align}
The inner sum on the left-hand side only goes to $n-1$, meaning that are no terms of $\phi_{m}$ on the right hand side. This adds a third condition on the energy, $E=\epsilon_m$, which proves Eq.~\eqref{eq:epseq}.

Finally, a physical meaning of the closed-form solutions can be obtained by Fourier-transforming equation \eqref{eq:3DFull},
\begin{equation}
    \frac{1}{2}|\mathbf{k}|^2\psi_\lambda(\mathbf{k})-\frac{4\pi}{6}\nabla^2\psi_\lambda(\mathbf{k})+4\pi s\frac{\psi_\lambda(\mathbf{k})}{2|\mathbf{k}|^2}=E\psi_\lambda(\mathbf{k}).
\end{equation}
By using the scaled coordinate $\mathbf{y}=\mathbf{k}/(4\pi)^\frac{1}{4}$ and multiplying both sides by $3$, we have
\begin{equation}
    -\frac{1}{2}\nabla^2\psi_\lambda(\mathbf{y})+\frac{3}{2}
    |\mathbf{y}|^2\psi_\lambda(\mathbf{y})+ 3s\frac{\psi_\lambda(\mathbf{y})}{2|\mathbf{y}|^2}=\Tilde{E}\psi_\lambda(\mathbf{y})
\end{equation}
where $\tilde{E}=\frac{3E}{\sqrt{4\pi}}$. This equation describes the isotropic harmonic oscillator at $\omega=\sqrt{3}$ with the exchange integral transformed into a centrifugal potential that increases the angular momentum in Fourier space. The analytical solutions appear at the values of $s$ for which the total angular momentum (kinetic plus exchange-induced) is integer, yielding back condition \eqref{eq:seq} for $\ell=0$. The fact that in the $\lambda\to\infty$ limit exchange contributes to the angular momentum in Fourier space also makes it clear why it enters at the same order ($\lambda^{-1/2}$) as the kinetic energy.

\section*{References}

\bibliographystyle{achemso}% Use the "unsrtnat" BibTeX style for formatting the Bibliography
\bibliography{bib_clean}

\end{document}